\documentclass[12pt,a4paper, table, xcdraw]{article}
\pdfoutput=1
\usepackage{geometry}
\UseRawInputEncoding
\usepackage{amsmath}
\usepackage{amsmath}
\usepackage{amssymb}
\usepackage[dvips]{graphicx}
\usepackage{cite}
\usepackage{pstricks}
\usepackage{bm}
\usepackage{bbm}
\usepackage{pbox}
\usepackage{placeins}
\usepackage{graphicx}
\usepackage{caption}
\usepackage{subcaption}
\usepackage[T1]{fontenc}
\usepackage{footnote}
\usepackage{pdfpages}
\usepackage{hhline}
\usepackage{multirow}
\usepackage{multicol}
\usepackage{enumitem}
\usepackage{multirow}
\usepackage{amsmath}
\usepackage{relsize}
 \usepackage{graphicx}
\usepackage[toc,page]{appendix}
\usepackage[english]{babel}
\allowdisplaybreaks
\usepackage{xcolor}
\usepackage{float}
\usepackage{slashed}
\usepackage{cases}
\usepackage{empheq}

\definecolor{MyDarkBlue}{rgb}{0.1, 0.3, 0.8} 
\definecolor{SBlue}{rgb}{0.2, 0.4, 0.4} 
\definecolor{MyLightBlue}{rgb}{0.22,0.51,0.99}
\definecolor{MyGreen}{rgb}{0.0, 0.5, 0.3}
\definecolor{BrickRed}{rgb}{0.8, 0.25, 0.33}
\usepackage[colorlinks=true,linkcolor=blue,citecolor=MyDarkBlue,
urlcolor=BrickRed,bookmarksnumbered=true,bookmarksopen]{hyperref}
\hypersetup{colorlinks, citecolor=SBlue,linkcolor=MyGreen, urlcolor=MyDarkBlue}
\begin{document}
\vspace*{-0.2in}
\begin{flushright}
\end{flushright}
\begin{center}
{\LARGE \bf
Flavor Seesaw Mechanism

}
\end{center}
\renewcommand{\thefootnote}{\fnsymbol{footnote}}
\begin{center}
{
{}~\textbf{Sudip Jana,}\footnote{ E-mail: \textcolor{MyDarkBlue}{sudip.jana@mpi-hd.mpg.de}}
{}~\textbf{Sophie Klett,}\footnote{ E-mail: \textcolor{MyDarkBlue}{sophie.klett@mpi-hd.mpg.de}}
{}~\textbf{Manfred Lindner}\footnote{ E-mail: \textcolor{MyDarkBlue}{lindner@mpi-hd.mpg.de}}
}
\vspace{0.5cm}
{
\\\em Max-Planck-Institut f{\"u}r Kernphysik, Saupfercheckweg 1, 69117 Heidelberg, Germany
} 
\end{center}
\renewcommand{\thefootnote}{\arabic{footnote}}
\setcounter{footnote}{0}
\thispagestyle{empty}

\begin{abstract}
In the Standard Model, Yukawa couplings parametrize the fermion masses and mixing angles with the  exception of  neutrino masses. The hierarchies and apparent regularities among the quark and lepton masses are, however, otherwise a mystery. 
We propose a new class of models having  vector-like fermions that can potentially address this problem and provide a new mechanism for fermion mass generation. The masses of the third and second generations of quarks and leptons arise at tree level via the seesaw mechanism from new physics at moderately higher scales, while loop corrections produce the masses for the first generation. This mechanism has a number of interesting and testable consequences. Among them are unavoidable flavor-violating signals at the upcoming experiments and the fact that neutrinos have naturally only Dirac masses.  
\end{abstract}
\newpage
\setcounter{footnote}{0}

\section{Introduction}\label{SEC-01}
The discovery of the Higgs boson in 2012 \cite{ATLAS:2012yve, CMS:2012qbp} completes the tremendous success of the Standard Model (SM). The SM has, however, a list of theoretical and experimental problems such that it cannot be the final theory. The observed masses and mixings of quarks and leptons do not belong to this list, even though the masses span an impressive 13 orders of magnitude from the light active neutrino mass scale to the top quark mass. The observed values are readily accommodated by the SM, albeit at the cost of adding Yukawa couplings with strengths ranging from $10^{-6}$ (for the electron) to $1$  (for the top quark). This is not a problem due to chiral symmetry which protects these drastically different values from big quantum corrections. The observed regularities of the masses and mixings remain, however, an unresolved mystery which may point to a mechanism beyond the SM which explains them. We present in this paper such a mechanism where new vector generations with TeV-ish masses lead to a seesaw like fermion mass matrix with tree level\footnote{For tree level realization, see Refs.~\cite{Berezhiani:1983hm,Chang:1986bp,Davidson:1987mh,Rajpoot:1987fca,Babu:1988mw,Babu:1989rb}.} and loop contributions.  Upon diagonalization the mechanism naturally produces the observed patterns without any extra ingredients.

The experimental data reveal that the quark mixing pattern differs significantly from the leptonic mixing pattern. The quark sector's mixing angles are tiny, meaning that the Cabibbo-Kobayashi-Maskawa (CKM) quark mixing matrix is quite close to the identity matrix. Two of the leptonic mixing angles, on the other hand, are large, and one is tiny, on the order of the Cabibbo angle, indicating a Pontecorvo-Maki-Nakagawa-Sakata (PMNS) leptonic mixing matrix that is substantially different from the identity matrix. This ``flavor puzzle'' is one of the features which the SM does not address. It provides motivation for investigating  models with enhanced field content and expanded flavour symmetry groups to explain the existing SM fermion mass spectrum and mixing parameters.

\begin{figure}[htb!]
		\centering
		\includegraphics[width=0.99\textwidth]{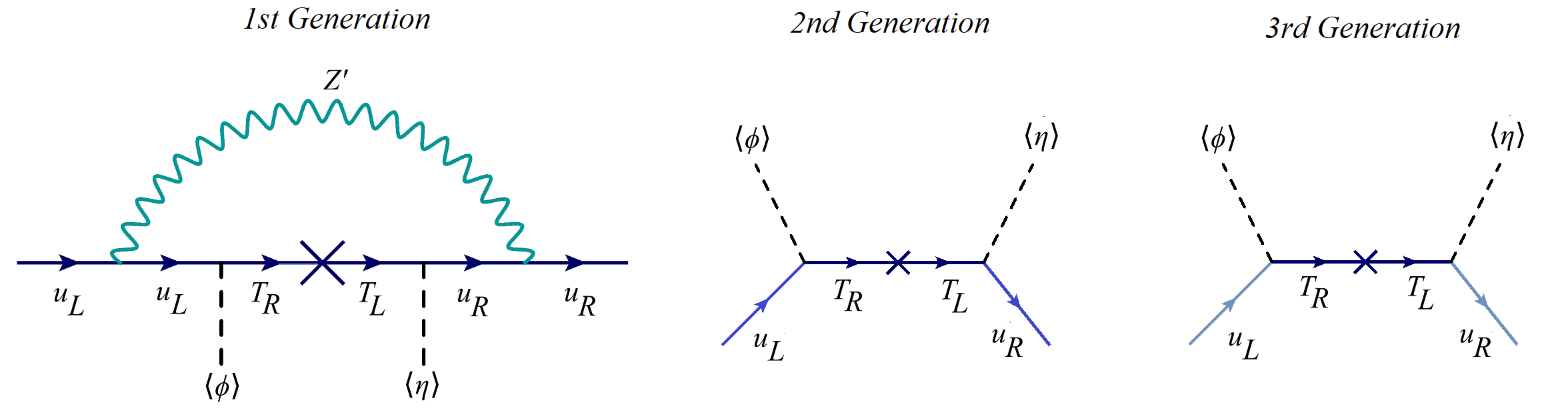}
	\caption{Schematic Feynman diagrams  for three generation quark and lepton masses.}
	\label{Fig:scheme}
\end{figure}

A primary step towards a solution of the ``flavor puzzle'' is to comprehend the physical characteristics of the generations. Partners from various generations have universal gauge interactions, but widely differing Yukawa couplings to the Higgs field. This may imply that there is  some underlying connection between gauge and Yukawa interactions while maintaining the cancellation of anomalies within a generation, which is a beautiful feature of the SM. 
The simplest gauge group with such qualities is widely known to be based on the difference of the baryon and lepton numbers, $U(1)_{B-L}$, provided a right-handed sterile neutrino per generation is added to cancel the  anomaly. The  $U(1)_{B-L}$ framework is widely studied in the literature in different contexts.  In the classical framework \cite{Pati:1974yy,Marshak:1979fm,Wilczek:1979et,Mohapatra:1980qe}, the $U(1)_{B-L}$ symmetry is considered to be broken at a very high scale ($\sim 10^{14}$ GeV) such that the tiny neutrino masses and mixings are generated via seesaw mechanism   by generating the lepton-number violating (LNV) Majorana mass for the sterile neutrino. The breaking of $U(1)_{B-L}$ symmetry at low  scale ($\sim$ TeV/ sub-TeV) has attracted quite a bit of interest recently \cite{Nelson:2007yq, Babu:2017olk}.
This class is substantially different from ours in terms of both philosophy and physics. Note also that one could identify the new abelian gauge group as a generic $U(1)_{X}$. Here,  the masses of the third and second generations of quarks and leptons arise in the tree level via the seesaw mechanism, while gauge loop corrections produce masses for the first generation (see  Fig.~\ref{Fig:scheme}). For a selection of models generating fermion mass hierarchies from quantum loop corrections see \cite{Balakrishna:1987qd,Babu:1988fn,Dobrescu:2008sz, Weinberg:2020zba, Graham:2009gr, Ma:2014yka}. In other theories, the mechanism for generating neutrino masses and mixing are, in general, quite different than the generation of quark and charged lepton masses. In contrast to that, our formalism is universally applicable to the up-type quarks, down-type quarks, charged leptons and neutrinos. It is  aesthetically  appealing to create such hierarchical mass pattern in a way that allows for natural Yukawa coupling [$\mathcal{O}(1)$] values as a result of loop suppression. Moreover, the neutrinos in most extensions of the standard model are presumed to be Majorana particles. The tiny neutrino masses and mixings are realized using the standard seesaw mechanism \cite{Minkowski:1977sc, Yanagida:1979as, GellMann:1980vs, Mohapatra:1979ia, Schechter:1980gr, Schechter:1981cv, Foot:1988aq}, which generally generates an effective dimension- 5 lepton number violating operator $\mathcal{O}_{1}=(L L H H) / \Lambda,$ suppressed by the mass scale $\Lambda$ of the heavy right-handed neutrino. However, if the lepton number is a conserved quantum number, Majorana neutrino masses are prohibited, and the standard seesaw mechanism does not work. It requires an alternative explanation for the smallness of the Dirac neutrino mass.  In models where the neutrino mass is zero at the tree level, it is possible to generate a small Dirac mass as a radiative correction. Such models are available both in the context of $\mathrm{SU}(2) \times U(1)$ gauge theory  with right-handed neutrinos and in the context of $\mathrm{SU}(2)_{L} \times \mathrm{SU}(2)_{R} \times \mathrm{U}(1)$ gauge theories. These models \cite{Mohapatra:1987hh, Mohapatra:1987nx, Balakrishna:1988bn, Branco:1978bz, Babu:1988yq, Gu:2007ug, Farzan:2012sa, Okada:2014vla, Bonilla:2016diq, Wang:2016lve, Ma:2017kgb, Wang:2017mcy, Helo:2018bgb, Reig:2018mdk, Han:2018zcn, Kang:2018lyy, Bonilla:2018ynb, Calle:2018ovc, Carvajal:2018ohk, Ma:2019yfo, Bolton:2019bou, Saad:2019bqf, Bonilla:2019hfb, Dasgupta:2019rmf, Jana:2019mez, Enomoto:2019mzl, Ma:2019byo, Jana:2019mgj, Restrepo:2019soi}, however, assume the existence of new fermions or bosons and in most cases new discrete symmetries as well, whose sole purpose is to provide an explanation for the small neutrino mass. We present a model of Dirac neutrino masses which does not suffer from this unsatisfactory feature. 

This paper is organized as follows: in the next section, we briefly discuss the basic mechanism. Then we describe in detail the proposed model, its symmetries, particle spectrum, gauge sector, scalar potential and Yukawa interactions. Subsequently, we analyze the resulting masses and mixings of quark and leptons and finally we discuss phenomenological implications before we conclude.

\section{The Flavor Seesaw Mechanism}\label{SEC-02}
Let us first sketch the main ingredients of the \textit{flavor seesaw} mechanism. Therefore consider the following mass matrix for the up quark sector, which has at the tree level the form:
$$
\left(\begin{array}{cc}
0 & v_{EW}|h\rangle \\
v_{S}\langle h| & M_{P}
\end{array}\right)~.
$$
Here  $|h \rangle$ is a $n$-element column vector including the Yukawa couplings,  $M_{P}$ 
the explicit mass  of a vector fermion
, $v_{EW}$ the electro-weak (EW) vacuum expectation value (VEV) and $v_{S}$ the VEV of a suitable new scalar. A model realizing these details will be presented in the next section. For the down sector, we have a similar matrix as well.  Note that the above matrix yields two nonzero and $n-1$ zero mass-eigenvalues. If $M_P>>v_{EW}, v_S$, which we call the seesaw limit, the nonzero eigenvalues are provided by
$$
m_{t} \simeq a_{0}\langle h \mid h\rangle, ~ m_{P} \simeq M_{P}
$$
where $a_{0}=-\left(v_{EW} v_{S} / M_{P}\right) .$ The eigenvector $|t\rangle$ corresponding to $m_{t}$ is proportional to $|h\rangle$. After including loop effects, we can write
$$
M_{T}=\left(\begin{array}{c|c}
\delta M & v_{EW}|\alpha\rangle \\
\hline v_{S}\langle\alpha| & M_{P}
\end{array}\right)
$$
where $\delta M$  originates from quantum loop effects and $|\alpha\rangle$ includes corrections. The rank of $M_{T} M_{T}^{\dagger}$ or $M_{T}^{\dagger} M_{T}$ which is also the rank of $M_{T}$ determines the number of massive fermions in the up sector. Counting the zero eigenvalues of $M_T$ yields its rank. We analyze the following eigenvalue equations in this regard.
$$
\begin{aligned}
\delta M|x\rangle+v_{EW}|\alpha\rangle x_{n+1} &=0 \\
v_{S}\langle\alpha \mid x\rangle+M_{P} x_{n+1} &=0
\end{aligned}
$$

By eliminating $x_{n+1}$, we obtain
$$
\left(\delta M+a_{0}|\alpha\rangle\langle\alpha|\right)|x\rangle \equiv M|x\rangle=0
$$
where we consider the combination $\delta M+a_{0}|\alpha\rangle\langle\alpha|$ as $M .$ If $M$ has rank $r$, then $M_{T} M_{T}^{\dagger}$ possesses $r+1$ nonzero eigenvalues and $r$ generations become massive. To investigate the mass hierarchy, we use the approach of calculating $M_T$ loop by loop and determining the number of massive generations based on the rank of the mass-matrix for the light quarks and leptons, $M$.
We discuss systematically scenarios with different numbers of vector-like fermion generations.

The above features are the basic mechanism which will be realized and phenomenologically analyzed in a specific model in the subsequent sections.

\section{Abelian Symmetry and the Flavor Puzzle}\label{SEC-03}

\subsection{Particle Spectrum}
In order to successfully implement the \textit{flavor seesaw} mechanism, we consider two additional generations of  singlet vector-like (VL)  fermions  for each species of SM fermions. More precisely, this includes VL up- and down-type quarks $T_k$, $B_k$, VL charged leptons $E_k$ and VL neutral leptons $N_k$, where $k$ is the generation index of these new particles. The gauge group of the model is given by $\mathcal{G}_{\mathrm{SM}} \times U(1)_{X}$ and a successful anomaly cancellation can be achieved by the addition of three generations of right handed neutrinos $\nu_{jR}$ with  $j = 1,2,3$. Finally, to break  $ U(1)_{X}$, a second scalar $\eta$ is required. The anomaly free charge assignments of all particles are given in Table~\ref{Tab.: Solution A particle content}. We emphasize that the SM Higgs is charged under $U(1)_{X}$ in this scenario and  the usual Higgs Yukawa couplings of the SM are forbidden. Note also that one could identify the new abelian gauge group $U(1)_{X}$ with $B-L$ charge.
\begin{table}[htb!]
		\centering
		\begin{tabular}{|c|c|c|c|c|}
			\hline \hline
			\bf Particle   & \bf{$\mathbf{SU(3)_C}$}  & \bf{$\mathbf{SU(2)_L}$} &\bf{$\mathbf{U(1)_Y}$} &\bf{$\mathbf{U(1)_{X}}$}\\
			\hline \hline
			$Q_{jL} = \left(
			\begin{array}{c}
				u_j\\
				d_j\\
			\end{array}
			\right)_L$  & \bf{3} &  \bf{2} &  1/3  & 1/3 \\
			$u_{j R}$   & \bf{3} &  \bf{1} &  4/3  & 1/3\\
			$d_{j R}$   & \bf{3} &  \bf{1} &  -2/3 & 1/3\\
			\hline
			$\Psi_{jL} = \left(
			\begin{array}{c}
			\nu_j\\
			e_j\\
			\end{array}
			\right)_L$  & \bf{1} &  \bf{2} &  -1   & -1\\
			$\nu_{j R}$ & \bf{1} &  \bf{1} &  0  & -1\\
			$e_{j R}$   & \bf{1} &  \bf{1} &  -2 & -1\\
			\hline
			$T_{k \, L}$, $T_{k \, R}$ & \bf{3} &  \bf{1} &  4/3  & 2/3\\
			$B_{k \, L}$, $B_{k \, R}$ & \bf{3} &  \bf{1} &  -2/3  & 0\\
			$N_{k \, L}$, $N_{k \, R}$ & \bf{1} &  \bf{1} &  0  & -2/3\\
			$E_{k \, L}$, $E_{k \, R}$ & \bf{1} &  \bf{1} &  -2  & -4/3\\
			\hline
			$\phi = \left(
			\begin{array}{c}
			\phi^{+}\\
			\phi^{0}\\
			\end{array}
			\right)$  & \bf{1} &  \bf{2} &  1   & 1/3\\
			$\eta$ & \bf{1} &  \bf{1} &  0  & 1/3\\
			\hline \hline
		\end{tabular}
		\caption{ Particle content and charge assignment under the gauge group $\mathcal{G}_{\mathrm{SM}} \times U(1)_{X}$, where $j = 1,2,3$ denotes the SM family and we introduce $k$ generations of  VL fermions. The electric charge is given by  $Q = T_3 + \frac{Y}{2}$, where $T_3$ is the third component of weak isospin .}
		\label{Tab.: Solution A particle content}
	\end{table}

\subsection{Gauge Boson Sector}
The Lagrangian of the gauge sector is given by
\begin{equation}
    \mathcal{L}_{\mathrm{gauge}} = -\dfrac{1}{4}G^a_{\mu \nu}G^{\mu \nu \, a}-\dfrac{1}{4}W^i_{\mu \nu}W^{\mu \nu \, i}-\dfrac{1}{4}B_{\mu \nu}B^{\mu \nu}-\dfrac{1}{4}X_{\mu \nu}X^{\mu \nu} \, ,
\end{equation}
where $G^a_{\mu \nu}, W^i_{\mu \nu}, B_{\mu \nu}$ and $X_{\mu \nu}$ denote the field strength tensors of the $SU(3)_C, SU(2)_L, U(1)_Y$ and $U(1)_{X}$ group respectively with  $a = 1,..., 8$ and $i = 1,..,3$ and we will indicate the couplings of these various gauge groups by $g_s$,  $g$, $g'$ and $g_X$.
Except for the VL down-type quark, all fermions are charged under $U(1)_{X}$ and will therefore interact with the neutral gauge boson $X$ that is associated with the new abelian gauge group. 
In general, the neutral current interaction involving the gauge boson $X$ is  given by \cite{Langacker:2000ju}
\begin{equation}
\label{Eq.:  NC Z' interaction basis}
    \mathcal{L}_{\mathrm{NC}} \supset \dfrac{g_X}{2} \sum_{i,j} \overline{\psi}_i \gamma_\mu \left[ q_\psi P_L + q_\psi P_R \right] \psi_j X^\mu \, ,
\end{equation}
where $\psi \in \{Q, \Psi, u, d, \nu, e, T, B,  N, E \}$ and $q_\psi$ denotes the respective charge under $U(1)_{X}$ (see Table~\ref{Tab.: Solution A particle content}). In the above equation we consider the most general case, where the couplings of $X$ are flavor non-diagonal. We will discuss the effects of these flavor changing neutral currents (FCNCs) in more detail in Sec.~\ref{SEC-04}. For now it is important to note that the generations within one fermion species are treated on an equal footing, i.e. their $U(1)_{X}$ charges are family independent. Moreover, left -and right-handed fields carry the same charges.

\subsection{Scalar Potential}
The additional scalar $\eta$ is a SM singlet charged under $U(1)_{X}$. It couples to the SM Higgs via the scalar portal term. Thus,
the Lagrangian in the scalar sector is given by 
\begin{equation}
\mathcal{L}_{\mathrm{scalar}} = (D_\mu \phi)^{\dagger}(D^\mu \phi) + (D_\mu \eta)^{\dagger}(D^\mu \eta) -V(\phi, \eta) ~,
\end{equation}
\begin{equation}
	V(\phi, \eta) = -\mu_{\phi}^2 \phi^{\dagger} \phi + \dfrac{1}{2}\lambda_{\phi} (\phi^{\dagger} \phi)^2  -\mu_{\eta}^2 \eta^{\dagger} \eta + \dfrac{1}{2}\lambda_{\eta} (\eta^{\dagger} \eta)^2 + \lambda_{\phi \eta} (\phi^{\dagger} \phi)(\eta^{\dagger} \eta) \, .
\end{equation}
In this notation $	V(\phi, \eta)$ describes the scalar potential and the covariant derivatives take the form
\begin{equation}
    D_\mu \phi = \left( \partial_\mu + i g\dfrac{\tau^i}{2}  W^{i}_\mu + i\dfrac{ g'}{2}B_\mu+ i q_{\phi}\dfrac{ g_X}{2}X_\mu\right)\phi \, ,
\end{equation}

\begin{equation}
    D_\mu \eta = \left( \partial_\mu + i q_{\eta} \dfrac{ g_X}{2}X_\mu\right)\eta \, ,
\end{equation}
with the $U(1)_{X}$ charge of the two scalars $q_\phi$ and $q_\eta$.
If the scalars acquire VEVs
\begin{equation}
	\langle\phi \rangle = \left(
	\begin{array}{c}
	0\\
	v_{EW}/ \sqrt{2}\\
	\end{array}
	\right) \, ,\indent \langle \eta \rangle = \dfrac{v_{S}}{ \sqrt{2}} \, ,
\end{equation}
 symmetry breaking occurs in two steps:
\begin{equation}
\begin{split}
		SU(3)_C \times SU(2)_L \times U(1)_Y \times U(1)_{X} &\overset{\langle\eta \rangle}{\longrightarrow} SU(3)_C \times SU(2)_L \times U(1)_Y \\
		&\overset{\langle\phi \rangle}{\longrightarrow} SU(3)_C \times   U(1)_{EM} ~,
\end{split}
\end{equation}
where we assume $v_{S} \gg v_{EW}$. As a result, the gauge bosons of the broken symmetries become massive.\\
Since $\eta$ is a weak singlet,  the electrically charged gauge bosons acquire mass solely due to the non-zero expectation value of $\phi$. Using the  convention $W_\mu^{\pm} = \left(W_\mu^2 \mp i W_\mu^1 \right)/\sqrt{2}$, their tree level mass is given by 
\begin{equation}
    M^2_{W^\pm} = \dfrac{g^2 v_{EW}^2}{4} \, .
\end{equation}
We note that this result is similar to the SM case. In the neutral gauge boson sector, mixing between the SM gauge bosons and the newly added $X$ is induced by $\phi$ carrying non-zero  charge under $U(1)_{X}$.
The mass matrix spanning the neutral gauge boson space $(B, W^3, X)$ has the form
\begin{equation}
    \mathcal{M}^2 = \dfrac{1}{4}\left(
	\begin{array}{c c c}
	g'^2 v_{EW}^2 & -g g' v_{EW}^2 &  g' g_X q_\phi v_{EW}^2 \\
	-g g'v_{EW}^2& g^2 v_{EW}^2&  -g \,  g_X q_\phi v_{EW}^2\\
	g' g_X q_\phi v_{EW}^2&  -g \, g_X q_\phi v_{EW}^2& g_X^2 \left(q^2_\phi v_{EW}^2 + q^2_\eta v_{S}^2 \right)\\
	\end{array}
	\right) \, .
\end{equation}
By defining the rotation angle $s_w \equiv \sin{\theta_w} = g' /\sqrt{g^2+g'^2}$ we can transform into a more convenient basis
\begin{equation}
\label{Eq.: Neutral Gauge Boson Mixing}
    \left(
    \begin{array}{c}
         A  \\
         Y \\
         X
    \end{array}
    \right)
    =
    \left(
    \begin{array}{ccc}
    c_w & s_w & 0 \\
    -s_w & c_w & 0\\
    0 & 0 & 1\\
    \end{array}
    \right)
    \left(
    \begin{array}{c}
         B  \\
         W^3\\
         X
    \end{array}
    \right) \, ,
\end{equation}
where the squared mass matrix  has one zero eigenvalue. This corresponds to the massless photon $A$. The submatrix mixing the two massive states $(Y, X)$ is then described by the entries
\begin{equation}
\label{Eq.: Neutral Gauge boson mass matrix undiagonalized}
    \begin{split}
        M^2_{YY} &= \dfrac{v_{EW}^2\left[ g^2 c^2_w (1+2s^2_w) + g'^2 s^2_w(1+2c^2_w)\right]}{4} \, ,\\ 
        M^2_{XX} &= \dfrac{g_X^2 (q^2_\phi v_{EW}^2 + q^2_\eta v_{S}^2)}{4} \, ,\\
         M^2_{YX} &= -\dfrac{g_X q_\phi v_{EW}^2(g c_w + g' s_w)}{4} \, .\\
    \end{split}
\end{equation}
Diagonalization by a further transformation 
\begin{equation}
    \left(
    \begin{array}{c}
         Z  \\
         Z'
    \end{array}
    \right)
    =
    \left(
    \begin{array}{cc}
    c_\xi & s_\xi  \\
    -s_\xi & c_\xi \\
    \end{array}
    \right)
    \left(
    \begin{array}{c}
         Y  \\
         X
    \end{array}
    \right) \, ,
\end{equation}
yields the two massive eigenstates $Z$ and $Z'$ with masses
\begin{equation}
    M^2_{Z, Z'} = \dfrac{1}{2} \left(M_{YY}^2 + M_{XX}^2 \mp  (M_{YY}^2 - M_{XX}^2)\sqrt{1+\tan^2{2\xi} } \right) \, ,
\end{equation}
and the mixing angle defined by 
\begin{equation}
\label{Eq.: Z-Z' mixing angle}
    \tan{2 \xi} = \dfrac{2M^2_{YX}}{M^2_{YY}-M^2_{XX}} \, .
\end{equation}
We note that although $Z-Z'$ mixing is induced in our model, the mixing angle will be suppressed by $\mathcal{O}(v^2_{EW}/v^2_S)$. Hence, in the limit $v_S \gg v_{EW} $, these effects are rather small and we will neglect them in the subsequent considerations.  In general, kinetic mixing may occur as well. However, we consider negligible kinetic mixing in our study.

\subsection{Yukawa Lagrangian}
Having specified the particle content and gauge interactions, it is particularly interesting to consider the explicit realization of the \textit{flavor seesaw}  in our model. As already noted, the charge assignments of the $U(1)_{X}$ gauge group forbid  Yukawa couplings between left -and right handed SM fermions via the usual Higgs mechanism. Instead, the following Yukawa couplings with the VL fermions are allowed
\begin{equation}
\label{Eq.: Yukawa Lagrangian}
\begin{split}
	\mathcal{L}_{\mathrm{Yuk}} = -&y^q_a  \overline{Q}_{jL} \tilde{\phi} T_{kR} - y^q_b \overline{T}_{kL} \eta \, u_{j R} -y^q_c  \overline{Q}_{jL} \phi B_{kR} - y^q_d \overline{B}_{kL} \eta^{\dagger} \, d_{j R} \\
	-&y^\ell_a  \overline{\Psi}_{jL} \tilde{\phi} N_{kR}- y^\ell_b \overline{N}_{kL} \eta \, \nu_{j R} 
	-y^\ell_c  \overline{\Psi}_{jL} \phi E_{kR} - y^\ell_d \overline{E}_{kL} \eta^{\dagger} \, e_{j R} + h.c. \, ,
\end{split}
\end{equation}
where $y_a,~y_b, ~y_c$ and $y_d$ denote the new Yukawa coupling matrices with  superscript $q$ ($\ell$) indicating the quark (lepton) sector.
Besides, the VL fermions can have explicit mass terms that are given by
\begin{equation}
    \mathcal{L}_{\mathrm{explicit}} = - \mathcal{M}_T \overline{T}_{kL} T_{kR} - \mathcal{M}_{B} \overline{B}_{kL} B_{kR} - \mathcal{M}_N \overline{N}_{kL} N_{kR}- \mathcal{M}_{E} \overline{E}_{kL} E_{kR} + h.c. \, .
\end{equation}
Without loss of generality we assume the VL mass matrices to be diagonal.

We like to emphasize that the assigned $U(1)_{X}$ charges prohibit direct Majorana mass terms for neutrinos. Moreover, within our minimal particle content there is no suitable scalar field which could generate a Majorana mass term from Yukawa interactions. In fact, the  LNV  Weinberg operator $LLHH/\Lambda$ is protected at any loop level, due to the $U(1)_X$ gauge symmetry.

\subsection{Generation of Quark Masses and Mixings}
\begin{figure}[ht!]
	\begin{subfigure}{0.5 \textwidth}
		\centering
		\includegraphics[scale=0.15]{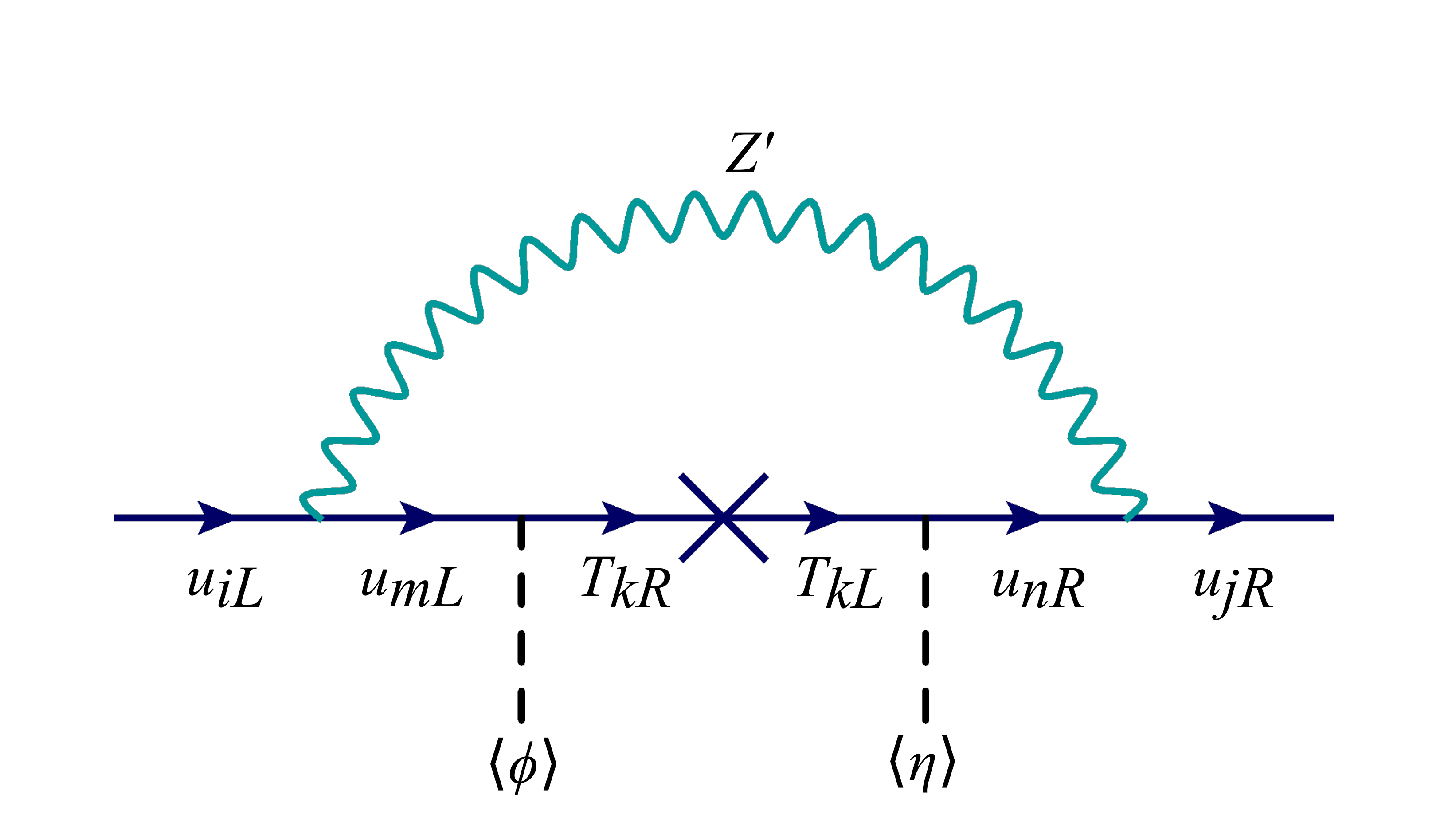}
		\caption{}
		\label{Fig.: Z' loop}
	\end{subfigure}
	\hfill
	\begin{subfigure}{0.5 \textwidth}
		\centering
		\includegraphics[scale=0.15]{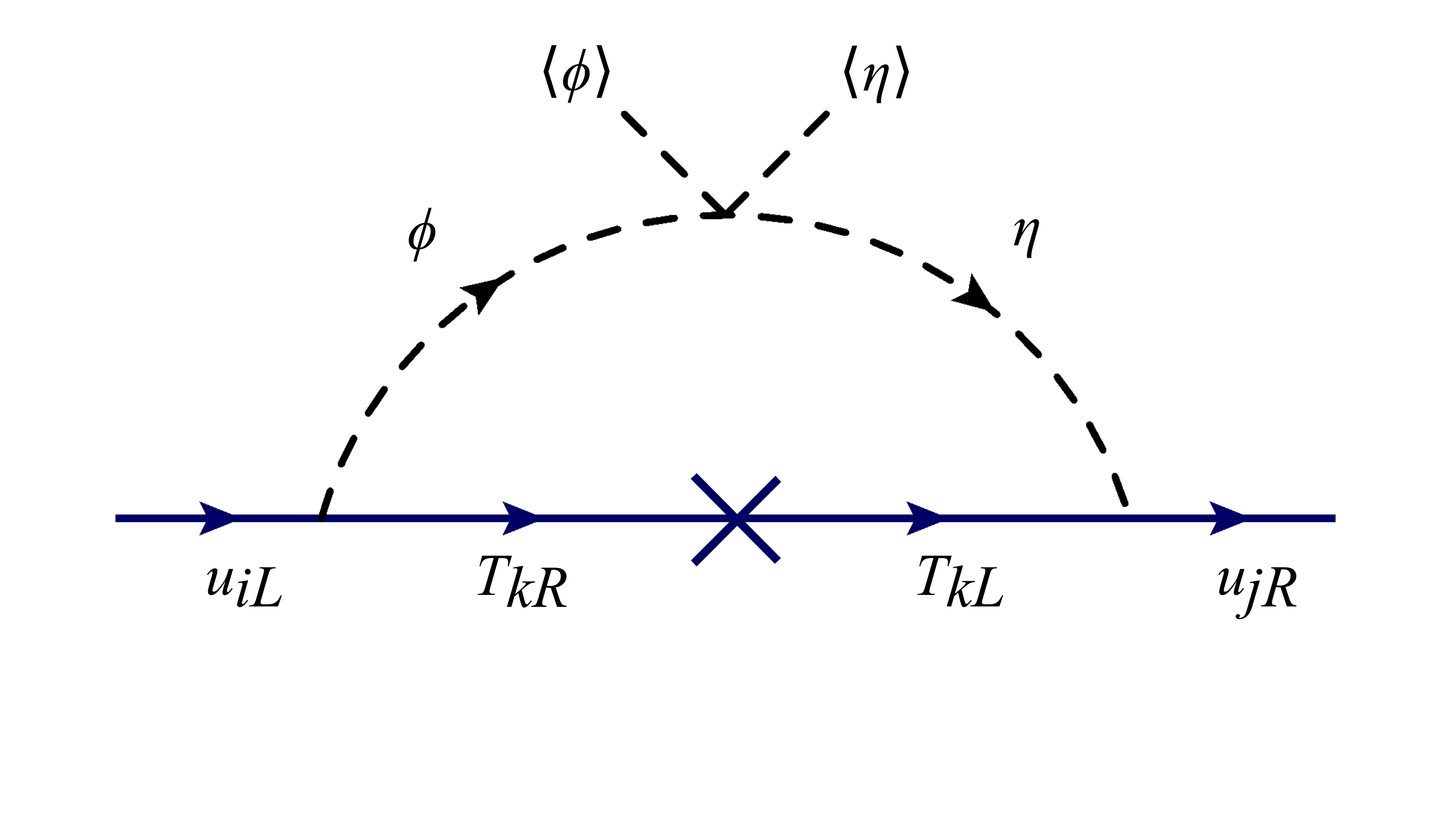}
		\caption{}
		\label{Fig.: Scalar Loop}
	\end{subfigure}
	\caption{\textbf{(a)} Feynman diagram  which contributes at 1-loop to the mass of the first generation quarks via $Z'$ boson exchange, where $i,j, m, n = 1,2,3$ denote the three generations of light quarks and $k$ labels the index of the VLF. \textbf{(b)} 1-loop contribution to the first generation masses via scalar exchange.}
	\label{Fig.: Loop contribution to Mass Matrix}
\end{figure} 
We start by considering the quark sector first.
After spontaneous symmetry breaking, the tree level mass matrix for the fermions receives contributions from both, the VEV of $\eta$ and $\phi$. In the up-type quark sector this can be written as 
\begin{equation}
    \overline{\mathbf{u}}_L \mathcal{M}^{(0)}_u \mathbf{u}_R\equiv
    \left(
	\begin{array}{ccccccc}
	\overline{u}_{1L} &
	\overline{u}_{2L}&
	\overline{u}_{3L}&
	\overline{T}_{1L}&
	\dots &
	\overline{T}_{kL}
	\end{array}
	\right)
	\left(
	\begin{array}{c|c}
	0_{3 \times3} & y^q_{a} \langle \phi \rangle  \\ \hline
	\left(y^q_{b}\right)^T \langle \eta \rangle & \mathcal{M}_T\\
	\end{array}
	\right)
	\left(
	\begin{array}{c}
	u_{1R} \\
	u_{2R} \\
	u_{3R} \\
	T_{1R}\\
	\dots \\
	T_{kR}\\
	\end{array}
	\right)
	\, ,
\end{equation}
with $y^q_a$  a $3 \times k$ matrix and likewise for all other Yukawa couplings.
Similarly, for the down-type quarks the mass matrix  is given by
\begin{equation}
     \overline{\mathbf{d}}_L\mathcal{M}^{(0)}_d\mathbf{d}_R\equiv 
     \left(
	\begin{array}{cccccc}
	\overline{d}_{1L} &
	\overline{d}_{2L} &
	\overline{d}_{3L} &
	\overline{B}_{1L}&
	\dots &
	\overline{B}_{kL}
	\end{array}
	\right)
    \left(
	\begin{array}{c|c}
	0_{3 \times3} & y^q_{c} \langle \phi \rangle  \\ \hline
	\left(y^q_{d}\right)^T \langle \eta \rangle & \mathcal{M}_B\\
	\end{array}
	\right)
	\left(
	\begin{array}{c}
	d_{1R} \\
	d_{2R} \\
	d_{3R} \\
	B_{1R}\\
	\dots \\
	B_{kR}\\
	\end{array}
	\right)
	\, .
\end{equation}
Provided that the column vectors of the respective Yukawa coupling matrices are linearly independent, the rank of these tree  level matrices is $2k$. Regarding the number of VL fermion generations, we categorize three different scenarios. First of all, consider the situation where the $Z'$ couplings in Eq.~(\ref{Eq.:  NC Z' interaction basis}) are flavor diagonal and thus the one-loop $Z'$ exchange contribution that is shown in Fig.~(\ref{Fig.: Z' loop}) is proportional to the tensor product of Yukawa couplings, i.e., for instance, $\sim y^q_a \left(y^q_b\right)^T$. By introducing three generations of VL fermions, all SM generations acquire masses at tree level, which is usually referred to as the universal seesaw mechanism \cite{Berezhiani:1983hm, Chang:1986bp, Davidson:1987mh, Rajpoot:1987fca, Babu:1988mw ,Babu:1989rb}. However, we find that with a proper choice of parameters the one-loop $Z'$ exchange diagram can be the dominant source of first-generation fermion masses. Secondly, one could also think of a scenario with four generations of VL fermions, of which one is massless and does not talk to the SM generations but only to the other VL fermions due to some proper symmetry. This implies that two generations of SM fermions become massive at tree level, while the first generation masses are generated from the one-loop $Z'$ exchange. In this work, we limit ourselves to a detailed discussion of a third case, where we consider flavor non-diagonal couplings of the $Z'$. Two massive generations of VL fermions can then generate tree-level masses for the third and second SM generation, while the one-loop contribution of the $Z'$ enhances the rank of the mass matrix by one, thus leading to first-generation masses. This can be understood in more detail considering the one-loop contribution of Fig.~(\ref{Fig.: Z' loop}) for non-diagonal $Z'$ couplings. 
In the following, we denote this contribution  by $\delta M^u_{ij} \equiv \Sigma_{ij}(\slashed{p}=0)$ which is in Landau gauge given by 
\begin{equation}
\label{Eq.: Loop Correction Z'}
\begin{split}
    \delta M^u_{ij} &= \sum_{k = 1}^2 \sum_{m,n = 1}^{3} i\dfrac{ 3 g_X^2 q_{Q} q_{u} [y^q_{a}]_{mk} [y^q_{b}]_{kn} v_{EW} v_{S}}{8}\int \dfrac{d^4 k}{(2 \pi)^4} \dfrac{\gamma_\mu (\slashed{k} +M_{Tk}) \gamma_\nu g^{\mu \nu}}{k^2 \left(k^2-M_{Tk}^2\right) \left((p-k)^2-M^2_{Z'}\right)} \\
     &= \sum_{k = 1}^2 \sum_{m,n = 1}^{3} \dfrac{ 3 g_X^2 q_{Q} q_{u} [y^q_{a}]_{mk} [y^q_{b}]_{kn} v_{EW} v_{S}}{32 \pi^2 } \dfrac{M_{Tk}}{\left( M_{Z'}^2-M_{Tk}^2\right)} \log{\dfrac{M_{Z'}^2}{M_{Tk}^2}} \, , 
\end{split}
\end{equation}
where we assume $k = 2$ from now on.
We note that the one-loop contribution of this diagram is independent of the index $i,j = 1, 2, 3$ as a consequence of non-diagonal $Z'$ couplings. Hence, the loop contribution is the same for all three generations of ordinary quarks and we can write $\delta M^u_{ij} \equiv \delta M^u$ in the subsequent.

Using this result, the mass matrix for the up-type sector is given to one-loop order by
\begin{equation}
\label{Eq.: Mass Matrix Up-type 1-loop}
     \overline{\mathbf{u}}_L \mathcal{M}^{(1)}_u \mathbf{u}_R\equiv
    \left(
	\begin{array}{cccccc}
	\overline{u}_{1L} &
	\overline{u}_{2L} &
	\overline{u}_{3L} &
	\overline{T}_{1L}&
	\overline{T}_{2L}&
	\end{array}
	\right)
		\left(
	\begin{array}{c|c}
	\mathcal{\delta M}^u & y^q_{a} \langle \phi \rangle  \\ \hline
	\left(y^q_{b}\right)^T \langle \eta \rangle & \mathcal{M}_T\\
	\end{array}
	\right)
	\left(
	\begin{array}{c}
	u_{1R} \\
	u_{2R} \\
	u_{3R} \\
	T_{1R}\\
	T_{2R}\\
	\end{array}
	\right)
	\, ,
\end{equation}
with  $\mathcal{\delta M}^u$ a $3 \times 3$ matrix where each entry equals $\delta M^u$. Starting from this general matrix form, it is evident that the column vectors of the Yukawa matrix $y^q_a$ need to be linearly independent from the column vectors of the loop contribution that are $\sim (1,1,1)^T$. Equal statements apply to all other Yukawa couplings in order to guarantee that $\mathcal{M}^{(1)}$ has rank five\footnote{
There exists a second diagram (see Fig.~\ref{Fig.: Scalar Loop}) which contributes at one-loop to the mass matrix. However we consider this diagram to be suppressed due to small scalar mixing. Furthermore its contribution is proportional to the tensor product $\sim y^q_a \left(y^q_b\right)^T$ and hence does not change the rank of $\mathcal{M}_u^{(1)}$.}.
It should be mentioned that there is an other one-loop contribution to the last column (row) of $\mathcal{M}_u^{(1)}$, since the VL fermions themselves couple to the $Z'$ (except for the bottom-type which carries no $U(1)_{X} $ charge). However, we regard this contribution to be small compared to the tree level entries and therefore neglect it in the further calculation.
We want to emphasize that the addition of two generations of VL fermions is the most minimal setup in order to generate masses for all three generations of SM  fermions. A further reduction of the number of vector generations would also reduce the rank of $\mathcal{M}^{(1)}$ and hence lead to one generation of SM fermions being massless. 

In the down-type sector, the one-loop mass matrix is similarly obtained and we quote here the result
\begin{equation}
    \label{Eq.: Mass Matrix down-type 1-loop}
     \overline{\mathbf{d}}_L \mathcal{M}^{(1)}_d \mathbf{d}_R\equiv
    \left(
	\begin{array}{cccccc}
	\overline{d}_{1L} &
	\overline{d}_{2L} &
	\overline{d}_{3L} &
	\overline{B}_{1L}&
	\overline{B}_{2L}&
	\end{array}
	\right)
		\left(
	\begin{array}{c|c}
	\mathcal{\delta M}^d & y^q_{c} \langle \phi \rangle  \\ \hline
	\left(y^q_{d}\right)^T \langle \eta \rangle & \mathcal{M}_T\\
	\end{array}
	\right)
	\left(
	\begin{array}{c}
	d_{1R} \\
	d_{2R} \\
	d_{3R} \\
	B_{1R}\\
	B_{2R}\\
	\end{array}
	\right)
	\, ,
\end{equation}
with the obvious replacements of the loop correction in Eq.~(\ref{Eq.: Loop Correction Z'}).
For the following, we diagonalize the obtained matrices by a bi-unitary transformation, such that 
\begin{equation}
\begin{split}
    V_L^u \mathcal{M}^{(1)}_u (V_R^u)^{\dagger} &=  \mathcal{M}^{\mathrm{diag}}_u \equiv \mathrm{diag}(m_u, m_c, m_t , m_{T1}, m_{T2}) \, ,\\
    V_L^d \mathcal{M}^{(1)}_d (V_R^d)^{\dagger} &=  \mathcal{M}^{\mathrm{diag}}_d \equiv \mathrm{diag}(m_d, m_s, m_b , m_{B1}, m_{B2}) \, .
    \end{split}
\end{equation}
Hence, the fermion mass eigenstates are given by
\begin{equation}
\label{Eq.: quark mass eigensate}
    \begin{split}
        \hat{\mathbf{u}}_{L/ R} = V^u_{L/R} \mathbf{u}_{L/R} \, ,\\
         \hat{\mathbf{d}}_{L/ R} = V^d_{L/R} \mathbf{d}_{L/R} \, .
    \end{split}
\end{equation}
Evidently, there is mixing between SM quarks and their VL partners, which will effect the gauge and Yukawa interactions when transforming to the fermion mass eigenbasis.

\subsection{Gauge and Yukawa  Interactions}
\label{Subsection: Gauge and Yukawa Interaction}
In the following section we illustrate our notation exemplary for the up-type sector.
In the flavor basis, the couplings of the up-type quarks to the $Z$ can be written as
\begin{equation}
    \mathcal{L} \supset Z_\mu \left[\overline{\mathbf{u}}_L \gamma^\mu g_L^u(Z) \mathbf{u}_L + \overline{\mathbf{u}}_R \gamma^\mu g_R^u(Z) \mathbf{u}_R \right] \, ,
\end{equation}
where the coupling strength is given by
\begin{equation}
\begin{split}
    g_L^u(Z) &= \dfrac{g}{c_w}\left[ \mathbbm{1} g_L^{u, SM}(Z) - \mathrm{diag}(0,0,0, \dfrac{1}{2}, \dfrac{1}{2}) \right] \, ,\\
    g_R^u(Z) &= \dfrac{g}{c_w}\left[ \mathbbm{1} g_R^{u, SM}(Z)  \right] \, ,
    \end{split}
\end{equation}
with $g_L^{u, SM}(Z) = T_3 - Q s^2_w= 1/2 -2/3 s^2_w$ and $g_R^{u, SM}(Z)= - Q s^2_w= -2/3 s^2_w$.
If we transform to the fermion mass basis, it is evident that there will be FCNC in the left-handed sector since $g_L^u(Z)$ is not proportional to the identity matrix and  the Lagrangian becomes
\begin{equation}
    \mathcal{L} \supset Z_\mu \left[\overline{\hat{\mathbf{u}}}_L \gamma^\mu \hat{g}_L^u(Z) \hat{\mathbf{u}}_L + \overline{\hat{\mathbf{u}}}_R \gamma^\mu \hat{g}_R^u(Z) \hat{\mathbf{u}}_R \right] \, , 
\end{equation}
with $\hat{g}_L^u(Z) \equiv  V_L^u g_L^u(Z) (V_L^u)^\dagger$ and $\hat{g}_R^u(Z) \equiv g_R^u(Z)$.\\
On an equal footing, we can write down the interaction with the $Z'$ boson  in the fermion mass eigenbasis
\begin{equation}
     \mathcal{L} \supset Z'_\mu \left[\overline{\hat{\mathbf{u}}}_L \gamma^\mu \hat{g}_L^u(Z') \hat{\mathbf{u}}_L + \overline{\hat{\mathbf{u}}}_R \gamma^\mu \hat{g}_R^{u}(Z') \hat{\mathbf{u}}_R \right] \, , 
\end{equation}
where 
\begin{equation}
    \begin{split}
        \hat{g}_L^u(Z') &= V_L^u \,g_L^u(Z') \,( V_L^u)^\dagger \, ,\\ 
        \hat{g}_R^u(Z') &= V_R^u \,g_R^u(Z') \,( V_R^u)^\dagger\, ,
    \end{split}
\end{equation}
and $g_L^u(Z') = g_R^u(Z')$ denote the coupling strengths in the interaction basis with non-diagonal entries proportional to the $U(1)_{X}$ charge of the fermion (compare Eq.~(\ref{Eq.:  NC Z' interaction basis})).  Consequently, there are tree level FCNCs mediated by both, the $Z$ and the $Z'$ boson. However, in the case of the $Z$ these effects are rather small since they are only induced by a small mixing between  SM quarks and  VL fermions.
   
In the mass basis, the coupling of the $W$ boson to quarks is given by 
\begin{equation}
    \mathcal{L} \supset \dfrac{g}{\sqrt{2}}W^+_\mu \left[\overline{\hat{\mathbf{u}}}_L \gamma^\mu \hat{g}_L^q(W) \hat{\mathbf{d}}_L  \right] + h.c. \, , 
\end{equation}
where 
\begin{equation}
    \begin{split}
        \hat{g}_L^q(W) &= V_L^u \,  g_L^q(W) \, ( V_L^d)^\dagger \, ,
    \end{split}
\end{equation}
and 
\begin{equation}
    \begin{split}
         g_L^q(W) &=  \mathrm{diag}(1,1,1,0, 0)\, .
    \end{split}
\end{equation}
From these definitions we find that the mixing matrix between the generations is described by a $5 \times 5 $ dimensional matrix, whose three dimensional submatrix yields the well-known CKM matrix
\begin{equation}
    V_{\mathrm{CKM}} \equiv \left. \hat{g}_L^q(W) \right\vert_{3 \times 3} \, .
\end{equation}
Finally, we consider the Yukawa couplings in the fermion mass eigenbasis.
In our model fermion masses receive not only contributions from the SM Higgs, but also from the second scalar $\eta$. Therefore, the couplings of fermions to the physical Higgs boson are no longer diagonal in the mass basis and give rise to tree level FCNC.
Adopting the same notation as before, the Higgs couplings in flavor basis from Eq.~(\ref{Eq.: Yukawa Lagrangian}) can be rewritten as
\begin{equation}
    \mathcal{L} \supset - \phi^0 \overline{\mathbf{u}}_L \mathbf{Y}_u \mathbf{u}_R 
    + h.c. \, , 
\end{equation}
where the Yukawa coupling matrix is given by
\begin{equation}
    \mathbf{Y}_u = \dfrac{1}{\sqrt{2}}
     \left(
	\begin{array}{cc}
	\mathbf{0}_{3x3} & y^q_a  \\
	\mathbf{0}_{2x3} & \mathbf{0}_{2x2}  \\
	\end{array}
	\right) \, \, .
\end{equation}
Transforming to the mass basis results in the non-diagonal coupling 
\begin{equation}
     \mathcal{L} \supset -\phi^0 \overline{\hat{\mathbf{u}}}_L \hat{\mathbf{Y}}_u \hat{\mathbf{u}}_R 
     + h.c.~,
\end{equation}
where we defined 
\begin{equation}
    \hat{\mathbf{Y}}_u = V^u_L \mathbf{Y}_u (V^u_R)^{\dagger} \,. 
\end{equation}
We note that the scalar $\eta$ also induces FCNC. However, since its mass can be at a high scale, we do not consider these effects here.

\subsection{Generation of Lepton Masses and Mixings}
The generation of neutrino masses succeeds analogously to the up-type quarks. At one-loop the mass term in the Lagrangian can be written as
\begin{equation}
    \overline{\mathbf{\bm{\nu}}}_L \mathcal{M}^{(1)}_{\nu} \mathbf{\bm{\nu}}_R\equiv
    \left(
	\begin{array}{cccccc}
	\overline{\nu}_{1L} &
	\overline{\nu}_{2L}&
	\overline{\nu}_{3L}&
	\overline{N}_{1L}&
	\overline{N}_{2L}&
	\end{array}
	\right)
	\left(
	\begin{array}{c|c}
	\delta \mathcal{M}^\nu & y^\ell_{a} \langle \phi \rangle  \\ \hline
	\left(y^\ell_{b}\right)^T \langle \eta \rangle & \mathcal{M}_N\\
	\end{array}
	\right)
	\left(
	\begin{array}{c}
	\nu_{1R} \\
	\nu_{2R} \\
	\nu_{3R} \\
	N_{1R}\\
	N_{2R}\\
	\end{array}
	\right)
	\, ,
\end{equation}
where $\delta \mathcal{M}^\nu$ is again a $3 \times 3$ matrix with each entry equal to $\delta M^\nu$. The loop contribution in the neutrino case can be inferred from Eq.~(\ref{Eq.: Loop Correction Z'}) by replacing the appropriate gauge charges, Yukawa couplings and VLF masses.
Analogously, for the charged leptons we find
\begin{equation}
    \overline{\mathbf{e}}_L \mathcal{M}^{(1)}_{e} \mathbf{e}_R\equiv
    \left(
	\begin{array}{cccccc}
	\overline{e}_{1L} &
	\overline{e}_{2L}&
	\overline{e}_{3L}&
	\overline{E}_{1L}&
	\overline{E}_{2L}&
	\end{array}
	\right)
	\left(
	\begin{array}{c|c}
	\delta \mathcal{M}^e & y^\ell_{c} \langle \phi \rangle  \\ \hline
	\left(y^\ell_{d}\right)^T \langle \eta \rangle & \mathcal{M}_E\\
	\end{array}
	\right)
	\left(
	\begin{array}{c}
	e_{1R} \\
	e_{2R} \\
	e_{3R} \\
	E_{1R}\\
	E_{2R}\\
	\end{array}
	\right)
	\, .
\end{equation}
These matrices can be diagonalized by the transformations
\begin{equation}
\begin{split}
V_L^\nu \mathcal{M}^{(1)}_\nu (V_R^\nu)^{\dagger} &=  \mathcal{M}^{\mathrm{diag}}_\nu \equiv \mathrm{diag}(m_{\nu_1}, m_{\nu_2}, m_{\nu_3} , m_{N1}, m_{N2}) \, , \\
    V_L^e \mathcal{M}^{(1)}_e (V_R^e)^{\dagger} &=  \mathcal{M}^{\mathrm{diag}}_e \equiv \mathrm{diag}(m_e, m_\mu, m_\tau , m_{E1}, m_{E2}) \, ,
    \end{split}
\end{equation}
where the fermion mass eigenstates are defined by
\begin{equation}
\label{Eq.: lepton mass eigensate}
    \begin{split}
     \hat{\bm{\nu}}_{L/ R} = V^\nu_{L/R} \bm{\nu}_{L/R} \, , \\
        \hat{\mathbf{e}}_{L/ R} = V^e_{L/R} \mathbf{e}_{L/R} \, .
    \end{split}
\end{equation}
Using these conventions, the charged current interaction in the mass basis reads
\begin{equation}
    \mathcal{L} \supset \dfrac{g}{\sqrt{2}}W^-_\mu \left[\overline{\hat{\mathbf{e}}}_L \gamma^\mu \hat{g}_L^\ell(W) \hat{\bm{\nu}}_L  \right] + h.c. \, , 
\end{equation}
where 
\begin{equation}
    \begin{split}
        \hat{g}_L^\ell(W) &= V_L^e \,  g_L^\ell(W) \, ( V_L^\nu)^\dagger \, ,
    \end{split}
\end{equation}
and 
\begin{equation}
    \begin{split}
         g_L^\ell(W) &=  \mathrm{diag}(1,1,1,0, 0)\, .
    \end{split}
\end{equation}
Following the standard convention, the PMNS matrix is then given by the $3 \times 3 $ submatrix:
\begin{equation}
    U_{\mathrm{PMNS}} \equiv \left. \hat{g}_L^\ell(W) \right\vert_{3 \times 3} \, .
\end{equation}
We note that in this scenario, the PMNS matrix is no longer unitary as assumed in the SM.
For the sake of brevity, we will not explicitly show here the neutral current interactions  but refer the reader to the conventions outlined in Sec.~\ref{Subsection: Gauge and Yukawa Interaction}.

\subsection{Numerical Results}
It is worth noting that the parameters for this type of flavor model cannot be 
chosen at random. The parameters of our model are mapped onto SM parameters and in
order to show that our model can reproduce all the observable perfectly, we give two benchmark points (BPs) and their predictions. The chosen Yukawa couplings in the quark and lepton sector are shown in Table~\ref{Tab:yukawabenchmarks},
\begin{table}[htb!]
		\centering 
		\resizebox{0.6\textwidth}{!}{%
			\begin{tabular}{|c|c|c|}
				\hline\hline
				\multirow{2}{*}{\begin{tabular}[c]{@{}c@{}}\bf{Yukawa Couplings}\\\end{tabular}} & \multicolumn{2}{c|}{\bf{Benchmark Points}}     \\ \cline{2-3} 
				&\bf{BP1}     & \bf{BP2}      \\ \hline \hline
				\rowcolor[HTML]{FFFDCA} 
\multicolumn{1}{|c|}{\cellcolor[HTML]{FFFDCA}
				$y^{q}_a$     }                                                              &  $\left(
				\begin{array}{cc}
					0.625 & 0.513\\
					0.186 \times e^{i 0.05} & 0.159\\
					0.401 & 0.327\\
				\end{array}
				\right)$      &      $\left(
				\begin{array}{cc}
					0.631 & 0.522\\
					0.183 \times e^{i 0.05} & 0.158\\
					0.412 & 0.344\\
				\end{array}
				\right)$   \\ \hline
					\rowcolor[HTML]{FFFDCA} 
\multicolumn{1}{|c|}{\cellcolor[HTML]{FFFDCA}
				$y^{q}_b$                    }                                               &  $\left(
				\begin{array}{cc}
					0.332 & 0.229\\
					0.340  & 0.224\\
					0.294 & 0.232 \times e^{i 0.05}\\
				\end{array}
				\right)$      &    $\left(
				\begin{array}{cc}
					0.335 & 0.227\\
					0.352  & 0.222\\
					0.290 & 0.230\times e^{i 0.05}\\
				\end{array}
				\right)$     \\ \hline
					\rowcolor[HTML]{FFFDCA} 
\multicolumn{1}{|c|}{\cellcolor[HTML]{FFFDCA}
				${y^{q}_c}$   }                                                                & $\left(
				\begin{array}{cc}
					1.539 &1.287\\
					0.195  & 0.538\\
					1.060 & 0.754 \\
				\end{array}
				\right)  \epsilon_1 $   &      $\left(
				\begin{array}{cc}
					1.501 & 1.341\\
					0.205  & 0.600\\
					1.061 & 0.795\\
				\end{array}
				\right) \epsilon_1 $  \\ \hline
					\rowcolor[HTML]{FFFDCA} 
\multicolumn{1}{|c|}{\cellcolor[HTML]{FFFDCA}
				$y^{q}_d$  }                                                                 & $\left(
				\begin{array}{cc}
					1.110 & 0.332\\
					0.850  & 1.506\\
				13.969 & 12.405 \\
				\end{array}
				\right)\epsilon_1 $       &    $\left(
				\begin{array}{cc}
					1.129 & 0.331\\
					0.811  & 1.624\\
					14.076 & 13.236\\
				\end{array}
				\right)\epsilon_1 $    \\ \hline\hline
					\rowcolor[HTML]{CCFACC} 
\multicolumn{1}{|c|}{\cellcolor[HTML]{CCFACC}
				$y^{\ell}_a$                 }                                                  &  $\left(
				\begin{array}{cc}
					1.306 & 1.494\\
					0.175 & 1.257\\
					0.538 & 0.333\\
				\end{array}
				\right) \epsilon_1^2$    &   $\left(
				\begin{array}{cc}
					1.235 \times e^{i 0.1} & 0.948 \\
					0.195 & 2.283\\
					0.443 & 0.347\\
				\end{array}
				\right) \epsilon_2$       \\ \hline
				\rowcolor[HTML]{CCFACC} 
\multicolumn{1}{|c|}{\cellcolor[HTML]{CCFACC}
				$y^{\ell}_b$         }                                                          &  $\left(
				\begin{array}{cc}
					0.319 & 1.048\\
				0.285 & 0.668\\
				0.967 & 0.328\\
				\end{array}
				\right)\epsilon_1^2 $    &    $\left(
				\begin{array}{cc}
					0.101 &1.210  \\
					2.871& 0.947\\
					0.243& 1.589\\
				\end{array}
				\right)\epsilon_2 $  \\ \hline
				\rowcolor[HTML]{CCFACC} 
\multicolumn{1}{|c|}{\cellcolor[HTML]{CCFACC}
				$y^{\ell}_c$                 }                                                  &  $\left(
				\begin{array}{cc}
					0.678 & 1.166\\
					0.854 & 0.574\\
					1.474 & 0.820\\
				\end{array}
				\right)\epsilon_1 $    &     $\left(
				\begin{array}{cc}
					0.484 & 1.468 \\
					0.999 & 1.281\\
					0.617 & 0.809\\
				\end{array}
				\right) \epsilon_1 $    \\ \hline
				\rowcolor[HTML]{CCFACC} 
\multicolumn{1}{|c|}{\cellcolor[HTML]{CCFACC}
				$y^{\ell}_d$                 }                                                  &  $\left(
				\begin{array}{cc}
					1.398 & 0.960\\
					0.740 & 0.780\\
					0.747 & 1.445\\
				\end{array}
				\right)\epsilon_1 $    &    $\left(
				\begin{array}{cc}
					1.555 & 0.479 \\
					1.355 & 1.381\\
					0.858& 0.982\\
				\end{array}
				\right) \epsilon_1 $   \\ \hline
				\hline
			\end{tabular}
		}
		\caption{Yukawa coupling benchmark points for our analysis.}
\label{Tab:yukawabenchmarks}
	\end{table}
 where $\epsilon_1 = 10^{-2}, \epsilon_2 = 10^{-1}\epsilon_1^2.$ For simplicity, we assume that the mass of the heavy gauge boson $M_Z'$ is $300$ TeV and the related gauge coupling $g_X$ equals 1 throughout our analysis. Considering the first benchmark point we select the VL quark masses $M_{T1} = 8.00 ~\mathrm{TeV}$ and $M_{B1} = 40.00 ~\mathrm{TeV}$ while the second generation is nearly degenerate with a mass difference  of $1$ GeV. In the lepton sector we use $M_{N1} = 7.00\times 10^7 ~\mathrm{TeV}$ and $M_{N2} = 1.00\times 10^8~\mathrm{TeV}$ for the VL neutral leptons and $ M_{E1} = 50.00 ~\mathrm{TeV}$ and $M_{E2} = 80.00~\mathrm{TeV}$ for charged VL leptons. For the second benchmark point we assume a larger degeneracy in the quark sector with  $M_{T1} = 8.00 ~\mathrm{TeV}$, $M_{T2} = 17.91 ~\mathrm{TeV}$, $M_{B1} =40.00 ~\mathrm{TeV}$ and $M_{B2} =65.69 ~\mathrm{TeV}$. The VL neutral lepton masses are given by $M_{N1} = 1.15\times 10^6 ~\mathrm{TeV}$ and $M_{N2} = 1.25\times 10^6 ~\mathrm{TeV}$ while those for the charged leptons are $M_{E1} =50.00 ~\mathrm{TeV}$ and $M_{E2} =80.00 ~\mathrm{TeV}$. Table~\ref{Tab.: Masses and Mixings with BPs} shows that these parameter settings are in accordance with the observable fermion masses and mixings.  Furthermore, as a measure of CP-violation, we calculated the  Jarlskog invariant which is defined by $\mathcal{J} \equiv \mathrm{Im}\left(V_{\mathrm{us}}V_{\mathrm{cb}}V^*_{\mathrm{ub}}V^*_{\mathrm{cs}} \right)$.
 
 Note that while the first benchmark point correctly reproduces a normal hierarchy (NH) for neutrino masses, the second benchmark point demonstrates that our model is also capable to describe an inverted hierarchy (IH). Using VL masses of a few tens to a few hundreds of TeV, the hierarchical mass structure of the quark sector (both up and down type) can be nicely accommodated for natural Yukawa coupling values of $\mathcal{O}(10^{-2})$-$\mathcal{O}(1)$. For the charged lepton sector  VL masses in the same range as in the quark sector also imply similarly moderate Yukawa couplings. For the neutral leptons VL masses of the order $\mathcal{O}(10^7)$ TeV would imply somewhat smaller Yukawa couplings $\mathcal{O}(10^{-4})$, which is still much less hierarchical than the usual values in the SM ranging from $10^{-13}$ to $1$. Note, however, that VL masses can have any value without introducing a new hierarchy problem. One could therefore choose higher values for the scale of the neutral lepton VL fermions and obtain in this way Yukawa couplings $\mathcal{O}(1)$. Choosing   $\mathcal{O}(1)$ Yukawa couplings would then require a VL mass scale of $\mathcal{O}(10^{17}~\mathrm{GeV})$ which leads via the seesaw formula to the correct neutrino masses. It is also important to keep in mind that the discussed benchmark points serve only as a proof of existence and a more detailed scan of the high dimensional parameter space is beyond the scope of this work. This implies that there could be further solutions with lower scales having more effects on a variety of BSM observable. Finally, it is important to stress that the number of parameters in our specific model does not allow to predict fermion masses or mixing angles. Instead the parameters are mapped in a way such that the observed hierarchies and mixing patterns emerge naturally. Other model realizations of the flavour seesaw mechanism may, however, be more restricted and thus be predictive. One can also imagine scenarios where symmetries among Yukawa couplings explain the observed hierarchies and mixing patterns in combination with the mapping.
\begin{table}[htb!]
\centering 
\resizebox{1.0\textwidth}{!}{%
\begin{tabular}{|cccccc|cccccc|}
\hline \hline
\multicolumn{6}{|c|}{\bf{Quark Sector}}                                                                                                                                                                                                                                                                 & \multicolumn{6}{c|}{\bf{Lepton Sector}}                                                                                                                                                                                                                                                                                                                                                                                                                                                                          \\ \hline
\multicolumn{1}{|c|}{}                                                                                          & \multicolumn{1}{c|}{}                                                  & \multicolumn{4}{c|}{\bf{Model Prediction}}                                                                   & \multicolumn{1}{c|}{}                                                                                                  & \multicolumn{1}{c|}{}                                                                            & \multicolumn{1}{c|}{}                                                                             & \multicolumn{1}{c|}{\begin{tabular}[c]{@{}c@{}}\bf{Model Prediction}\\ \bf{(NH)}\end{tabular}} & \multicolumn{2}{c|}{\begin{tabular}[c]{@{}c@{}}\bf{Model Prediction}\\ \bf{(IH)}\end{tabular}} \\ \cline{3-6} \cline{10-12} 
\multicolumn{1}{|c|}{\multirow{-2}{*}{\begin{tabular}[c]{@{}c@{}}\bf{Observable}\\  \bf{(Masses in GeV)}\end{tabular}}} & \multicolumn{1}{c|}{\multirow{-2}{*}{\bf{Exp. Range}}}                      & \multicolumn{2}{c|}{\bf{BP1}}                           & \multicolumn{2}{c|}{\bf{BP2}}                           & \multicolumn{1}{c|}{\multirow{-2}{*}{\begin{tabular}[c]{@{}c@{}}\bf{Observable}\\ \bf{(Masses in GeV)}\end{tabular}}}         & \multicolumn{1}{c|}{\multirow{-2}{*}{\begin{tabular}[c]{@{}c@{}}\bf{Exp. Range}\\ \bf{(NH)}\end{tabular}}} & \multicolumn{1}{c|}{\multirow{-2}{*}{\begin{tabular}[c]{@{}c@{}}\bf{Exp. Range} \\ \bf{(IH)}\end{tabular}}} & \multicolumn{1}{c|}{\bf{BP1}}                                                             & \multicolumn{2}{c|}{\bf{BP2}}                                                             \\ \hline \hline
\rowcolor[HTML]{FCE4C6} 
\multicolumn{1}{|c|}{\cellcolor[HTML]{FCE4C6}$m_u / 10^{-3}$}                                                   & \multicolumn{1}{c|}{\cellcolor[HTML]{FCE4C6}$1.38 \rightarrow 3.63$}   & \multicolumn{2}{c|}{\cellcolor[HTML]{FCE4C6}2.12}  & \multicolumn{2}{c|}{\cellcolor[HTML]{FCE4C6}3.07}  & \multicolumn{1}{c|}{\cellcolor[HTML]{FCE4C6}$\dfrac{\Delta m_{21}^2 }{ 10^{-5}\, \mathrm{eV}^2}$}                      & \multicolumn{1}{c|}{\cellcolor[HTML]{FCE4C6}$6.82 \rightarrow 8.04$}                             & \multicolumn{1}{c|}{\cellcolor[HTML]{FCE4C6}$6.82 \rightarrow 8.04$}                              & \multicolumn{1}{c|}{\cellcolor[HTML]{FCE4C6}7.583}                                   & \multicolumn{2}{c|}{\cellcolor[HTML]{FCE4C6}7.898}                                   \\ \hline
\rowcolor[HTML]{FCE4C6} 
\multicolumn{1}{|c|}{\cellcolor[HTML]{FCE4C6}$m_c$}                                                             & \multicolumn{1}{c|}{\cellcolor[HTML]{FCE4C6}$1.21 \rightarrow 1.33$}   & \multicolumn{2}{c|}{\cellcolor[HTML]{FCE4C6}1.29}  & \multicolumn{2}{c|}{\cellcolor[HTML]{FCE4C6}1.25}  & \multicolumn{1}{c|}{\cellcolor[HTML]{FCE4C6}}                                                                          & \multicolumn{1}{c|}{\cellcolor[HTML]{FCE4C6}}                                                    & \multicolumn{1}{c|}{\cellcolor[HTML]{FCE4C6}}                                                     & \multicolumn{1}{c|}{\cellcolor[HTML]{FCE4C6}}                                        & \multicolumn{2}{c|}{\cellcolor[HTML]{FCE4C6}}                                        \\ \cline{1-6}
\rowcolor[HTML]{FCE4C6} 
\multicolumn{1}{|c|}{\cellcolor[HTML]{FCE4C6}$m_t$}                                                             & \multicolumn{1}{c|}{\cellcolor[HTML]{FCE4C6}$171.7 \rightarrow 174.1$} & \multicolumn{2}{c|}{\cellcolor[HTML]{FCE4C6}172.3} & \multicolumn{2}{c|}{\cellcolor[HTML]{FCE4C6}174.1} & \multicolumn{1}{c|}{\multirow{-2}{*}{\cellcolor[HTML]{FCE4C6}$\dfrac{\Delta m_{3\ell}^2 }{ 10^{-3}\, \mathrm{eV}^2}$}} & \multicolumn{1}{c|}{\multirow{-2}{*}{\cellcolor[HTML]{FCE4C6}$2.421 \rightarrow 2.598$}}         & \multicolumn{1}{c|}{\multirow{-2}{*}{\cellcolor[HTML]{FCE4C6}$-2.583 \rightarrow -2.412$}}        & \multicolumn{1}{c|}{\multirow{-2}{*}{\cellcolor[HTML]{FCE4C6}2.567}}                 & \multicolumn{2}{c|}{\multirow{-2}{*}{\cellcolor[HTML]{FCE4C6}$-2.432$}}              \\ \hline
\rowcolor[HTML]{FFFDCA} 
\multicolumn{1}{|c|}{\cellcolor[HTML]{FFFDCA}$m_d/ 10^{-3} $}                                                   & \multicolumn{1}{c|}{\cellcolor[HTML]{FFFDCA}$4.16 \rightarrow 6.11$}   & \multicolumn{2}{c|}{\cellcolor[HTML]{FFFDCA}4.34}  & \multicolumn{2}{c|}{\cellcolor[HTML]{FFFDCA}5.08}  & \multicolumn{1}{c|}{\cellcolor[HTML]{FFFDCA}$m_e / 10^{-3}$}                                                           & \multicolumn{2}{c|}{\cellcolor[HTML]{FFFDCA}$0.485 \rightarrow 0.537$}                                                                                                                               & \multicolumn{1}{c|}{\cellcolor[HTML]{FFFDCA}0.511}                                   & \multicolumn{2}{c|}{\cellcolor[HTML]{FFFDCA}0.527}                                   \\ \hline
\rowcolor[HTML]{FFFDCA} 
\multicolumn{1}{|c|}{\cellcolor[HTML]{FFFDCA}$m_s $}                                                            & \multicolumn{1}{c|}{\cellcolor[HTML]{FFFDCA}$0.078 \rightarrow 0.126$} & \multicolumn{2}{c|}{\cellcolor[HTML]{FFFDCA}0.122} & \multicolumn{2}{c|}{\cellcolor[HTML]{FFFDCA}0.109} & \multicolumn{1}{c|}{\cellcolor[HTML]{FFFDCA}$m_{\mu}$}                                                                 & \multicolumn{2}{c|}{\cellcolor[HTML]{FFFDCA}$0.100 \rightarrow 0.111$}                                                                                                                               & \multicolumn{1}{c|}{\cellcolor[HTML]{FFFDCA}0.109}                                   & \multicolumn{2}{c|}{\cellcolor[HTML]{FFFDCA}0.109}                                   \\ \hline
\rowcolor[HTML]{FFFDCA} 
\multicolumn{1}{|c|}{\cellcolor[HTML]{FFFDCA}$m_b$}                                                             & \multicolumn{1}{c|}{\cellcolor[HTML]{FFFDCA}$4.12 \rightarrow 4.27$}   & \multicolumn{2}{c|}{\cellcolor[HTML]{FFFDCA}4.18}  & \multicolumn{2}{c|}{\cellcolor[HTML]{FFFDCA}4.13}  & \multicolumn{1}{c|}{\cellcolor[HTML]{FFFDCA}$m_{\tau}$}                                                                & \multicolumn{2}{c|}{\cellcolor[HTML]{FFFDCA}$1.688 \rightarrow 1.866$}                                                                                                                               & \multicolumn{1}{c|}{\cellcolor[HTML]{FFFDCA}1.862}                                   & \multicolumn{2}{c|}{\cellcolor[HTML]{FFFDCA}1.839}                                   \\ \hline
\rowcolor[HTML]{FFE2E0} 
\multicolumn{1}{|c|}{\cellcolor[HTML]{FFE2E0}$|V_{\mathrm{ud}}|$}                                               & \multicolumn{1}{c|}{\cellcolor[HTML]{FFE2E0}$0.973 \rightarrow 0.974$} & \multicolumn{2}{c|}{\cellcolor[HTML]{FFE2E0}0.974} & \multicolumn{2}{c|}{\cellcolor[HTML]{FFE2E0}0.974} & \multicolumn{1}{c|}{\cellcolor[HTML]{FFE2E0}}                                                                          & \multicolumn{1}{c|}{\cellcolor[HTML]{FFE2E0}}                                                    & \multicolumn{1}{c|}{\cellcolor[HTML]{FFE2E0}}                                                     & \multicolumn{1}{c|}{\cellcolor[HTML]{FFE2E0}}                                        & \multicolumn{2}{c|}{\cellcolor[HTML]{FFE2E0}}                                        \\ \cline{1-6}
\rowcolor[HTML]{FFE2E0} 
\multicolumn{1}{|c|}{\cellcolor[HTML]{FFE2E0}$|V_{\mathrm{us}}|$}                                               & \multicolumn{1}{c|}{\cellcolor[HTML]{FFE2E0}$0.222 \rightarrow 0.227$} & \multicolumn{2}{c|}{\cellcolor[HTML]{FFE2E0}0.227} & \multicolumn{2}{c|}{\cellcolor[HTML]{FFE2E0}0.226} & \multicolumn{1}{c|}{\cellcolor[HTML]{FFE2E0}}                                                                          & \multicolumn{1}{c|}{\cellcolor[HTML]{FFE2E0}}                                                    & \multicolumn{1}{c|}{\cellcolor[HTML]{FFE2E0}}                                                     & \multicolumn{1}{c|}{\cellcolor[HTML]{FFE2E0}}                                        & \multicolumn{2}{c|}{\cellcolor[HTML]{FFE2E0}}                                        \\ \cline{1-6}
\rowcolor[HTML]{FFE2E0} 
\multicolumn{1}{|c|}{\cellcolor[HTML]{FFE2E0}$|V_{\mathrm{ub}}| /10^{-4}$}                                      & \multicolumn{1}{c|}{\cellcolor[HTML]{FFE2E0}$31.0\rightarrow 45.4$}    & \multicolumn{2}{c|}{\cellcolor[HTML]{FFE2E0}38.4}  & \multicolumn{2}{c|}{\cellcolor[HTML]{FFE2E0}44.8}  & \multicolumn{1}{c|}{\multirow{-3}{*}{\cellcolor[HTML]{FFE2E0}$\sin^2(\theta_{12})$}}                                   & \multicolumn{1}{c|}{\multirow{-3}{*}{\cellcolor[HTML]{FFE2E0}$0.269 \rightarrow 0.343$}}         & \multicolumn{1}{c|}{\multirow{-3}{*}{\cellcolor[HTML]{FFE2E0}$0.269 \rightarrow 0.343$}}          & \multicolumn{1}{c|}{\multirow{-3}{*}{\cellcolor[HTML]{FFE2E0}0.315}}                 & \multicolumn{2}{c|}{\multirow{-3}{*}{\cellcolor[HTML]{FFE2E0}0.320}}                 \\ \hline
\rowcolor[HTML]{FFE2E0} 
\multicolumn{1}{|c|}{\cellcolor[HTML]{FFE2E0}$|V_{\mathrm{cd}}|$}                                               & \multicolumn{1}{c|}{\cellcolor[HTML]{FFE2E0}$0.209 \rightarrow 0.233$} & \multicolumn{2}{c|}{\cellcolor[HTML]{FFE2E0}0.226} & \multicolumn{2}{c|}{\cellcolor[HTML]{FFE2E0}0.226} & \multicolumn{1}{c|}{\cellcolor[HTML]{FFE2E0}}                                                                          & \multicolumn{1}{c|}{\cellcolor[HTML]{FFE2E0}}                                                    & \multicolumn{1}{c|}{\cellcolor[HTML]{FFE2E0}}                                                     & \multicolumn{1}{c|}{\cellcolor[HTML]{FFE2E0}}                                        & \multicolumn{2}{c|}{\cellcolor[HTML]{FFE2E0}}                                        \\ \cline{1-6}
\rowcolor[HTML]{FFE2E0} 
\multicolumn{1}{|c|}{\cellcolor[HTML]{FFE2E0}$|V_{\mathrm{cs}}|$}                                               & \multicolumn{1}{c|}{\cellcolor[HTML]{FFE2E0}$0.954 \rightarrow 1.020$} & \multicolumn{2}{c|}{\cellcolor[HTML]{FFE2E0}0.973} & \multicolumn{2}{c|}{\cellcolor[HTML]{FFE2E0}0.973} & \multicolumn{1}{c|}{\cellcolor[HTML]{FFE2E0}}                                                                          & \multicolumn{1}{c|}{\cellcolor[HTML]{FFE2E0}}                                                    & \multicolumn{1}{c|}{\cellcolor[HTML]{FFE2E0}}                                                     & \multicolumn{1}{c|}{\cellcolor[HTML]{FFE2E0}}                                        & \multicolumn{2}{c|}{\cellcolor[HTML]{FFE2E0}}                                        \\ \cline{1-6}
\rowcolor[HTML]{FFE2E0} 
\multicolumn{1}{|c|}{\cellcolor[HTML]{FFE2E0}$|V_{\mathrm{cb}}| / 10^{-3}$}                                     & \multicolumn{1}{c|}{\cellcolor[HTML]{FFE2E0}$36.8 \rightarrow 45.2$}   & \multicolumn{2}{c|}{\cellcolor[HTML]{FFE2E0}42.3}  & \multicolumn{2}{c|}{\cellcolor[HTML]{FFE2E0}41.9}  & \multicolumn{1}{c|}{\multirow{-3}{*}{\cellcolor[HTML]{FFE2E0}$\sin^2(\theta_{23})$}}                                   & \multicolumn{1}{c|}{\multirow{-3}{*}{\cellcolor[HTML]{FFE2E0}$0.407 \rightarrow 0.618$}}         & \multicolumn{1}{c|}{\multirow{-3}{*}{\cellcolor[HTML]{FFE2E0}$0.411 \rightarrow 0.621$}}          & \multicolumn{1}{c|}{\multirow{-3}{*}{\cellcolor[HTML]{FFE2E0}0.444}}                 & \multicolumn{2}{c|}{\multirow{-3}{*}{\cellcolor[HTML]{FFE2E0}0.413}}                 \\ \hline
\rowcolor[HTML]{FFE2E0} 
\multicolumn{1}{|c|}{\cellcolor[HTML]{FFE2E0}$|V_{\mathrm{td}}| / 10^{-4}$}                                     & \multicolumn{1}{c|}{\cellcolor[HTML]{FFE2E0}$71.0 \rightarrow 89.0$}   & \multicolumn{2}{c|}{\cellcolor[HTML]{FFE2E0}84.0}  & \multicolumn{2}{c|}{\cellcolor[HTML]{FFE2E0}78.7}  & \multicolumn{1}{c|}{\cellcolor[HTML]{FFE2E0}}                                                                          & \multicolumn{1}{c|}{\cellcolor[HTML]{FFE2E0}}                                                    & \multicolumn{1}{c|}{\cellcolor[HTML]{FFE2E0}}                                                     & \multicolumn{1}{c|}{\cellcolor[HTML]{FFE2E0}}                                        & \multicolumn{2}{c|}{\cellcolor[HTML]{FFE2E0}}                                        \\ \cline{1-6}
\rowcolor[HTML]{FFE2E0} 
\multicolumn{1}{|c|}{\cellcolor[HTML]{FFE2E0}$|V_{\mathrm{ts}}| /10^{-3}$}                                      & \multicolumn{1}{c|}{\cellcolor[HTML]{FFE2E0}$35.5 \rightarrow 42.1$}   & \multicolumn{2}{c|}{\cellcolor[HTML]{FFE2E0}41.6}  & \multicolumn{2}{c|}{\cellcolor[HTML]{FFE2E0}41.4}  & \multicolumn{1}{c|}{\cellcolor[HTML]{FFE2E0}}                                                                          & \multicolumn{1}{c|}{\cellcolor[HTML]{FFE2E0}}                                                    & \multicolumn{1}{c|}{\cellcolor[HTML]{FFE2E0}}                                                     & \multicolumn{1}{c|}{\cellcolor[HTML]{FFE2E0}}                                        & \multicolumn{2}{c|}{\cellcolor[HTML]{FFE2E0}}                                        \\ \cline{1-6}
\rowcolor[HTML]{FFE2E0} 
\multicolumn{1}{|c|}{\cellcolor[HTML]{FFE2E0}$|V_{\mathrm{tb}}|$}                                               & \multicolumn{1}{c|}{\cellcolor[HTML]{FFE2E0}$0.923 \rightarrow 1.103$} & \multicolumn{2}{c|}{\cellcolor[HTML]{FFE2E0}0.999} & \multicolumn{2}{c|}{\cellcolor[HTML]{FFE2E0}0.999} & \multicolumn{1}{c|}{\multirow{-3}{*}{\cellcolor[HTML]{FFE2E0}$\sin^2(\theta_{13})$}}                                   & \multicolumn{1}{c|}{\multirow{-3}{*}{\cellcolor[HTML]{FFE2E0}$0.02034 \rightarrow 0.02430$}}     & \multicolumn{1}{c|}{\multirow{-3}{*}{\cellcolor[HTML]{FFE2E0}$0.02053 \rightarrow 0.02436$}}      & \multicolumn{1}{c|}{\multirow{-3}{*}{\cellcolor[HTML]{FFE2E0}0.02053}}                 & \multicolumn{2}{c|}{\multirow{-3}{*}{\cellcolor[HTML]{FFE2E0}0.02300}}                 \\ \hline
\rowcolor[HTML]{CCFACC} 
\multicolumn{1}{|c|}{\cellcolor[HTML]{CCFACC}$\mathcal{J}/ 10^{-5}$}                                            & \multicolumn{1}{c|}{\cellcolor[HTML]{CCFACC}$2.73 \rightarrow 3.45$}   & \multicolumn{2}{c|}{\cellcolor[HTML]{CCFACC}3.12}  & \multicolumn{2}{c|}{\cellcolor[HTML]{CCFACC}3.40}  & \multicolumn{1}{c|}{\cellcolor[HTML]{CCFACC}$\delta_{\mathrm{cp} }/\circ$}                                             & \multicolumn{1}{c|}{\cellcolor[HTML]{CCFACC}$107 \rightarrow 403$}                               & \multicolumn{1}{c|}{\cellcolor[HTML]{CCFACC}$192 \rightarrow 360$}                                & \multicolumn{1}{c|}{\cellcolor[HTML]{CCFACC}0}                                       & \multicolumn{2}{c|}{\cellcolor[HTML]{CCFACC}250}                                     \\ \hline \hline
\end{tabular}
}
\caption{SM observable for the two benchmark points. Here $\Delta m^2_{3\ell} \equiv \Delta m^2_{31}$ for NH and $\Delta m^2_{3\ell} \equiv \Delta m^2_{32}$ for IH. Experimental ranges \cite{Zyla:2020zbs,Esteban:2020cvm} denote $3\sigma$ intervals except for the charged lepton masses where we demand our model to fulfil the measured value within $\pm  5\%$. }
\label{Tab.: Masses and Mixings with BPs}
\end{table}

\section{Phenomenological Implications}
\label{SEC-04}
\subsection{FCNC Processes in Quark Sector} 
\begin{figure}[ht]
	\begin{subfigure}{0.5 \textwidth}
		\centering
		\includegraphics[scale=0.15]{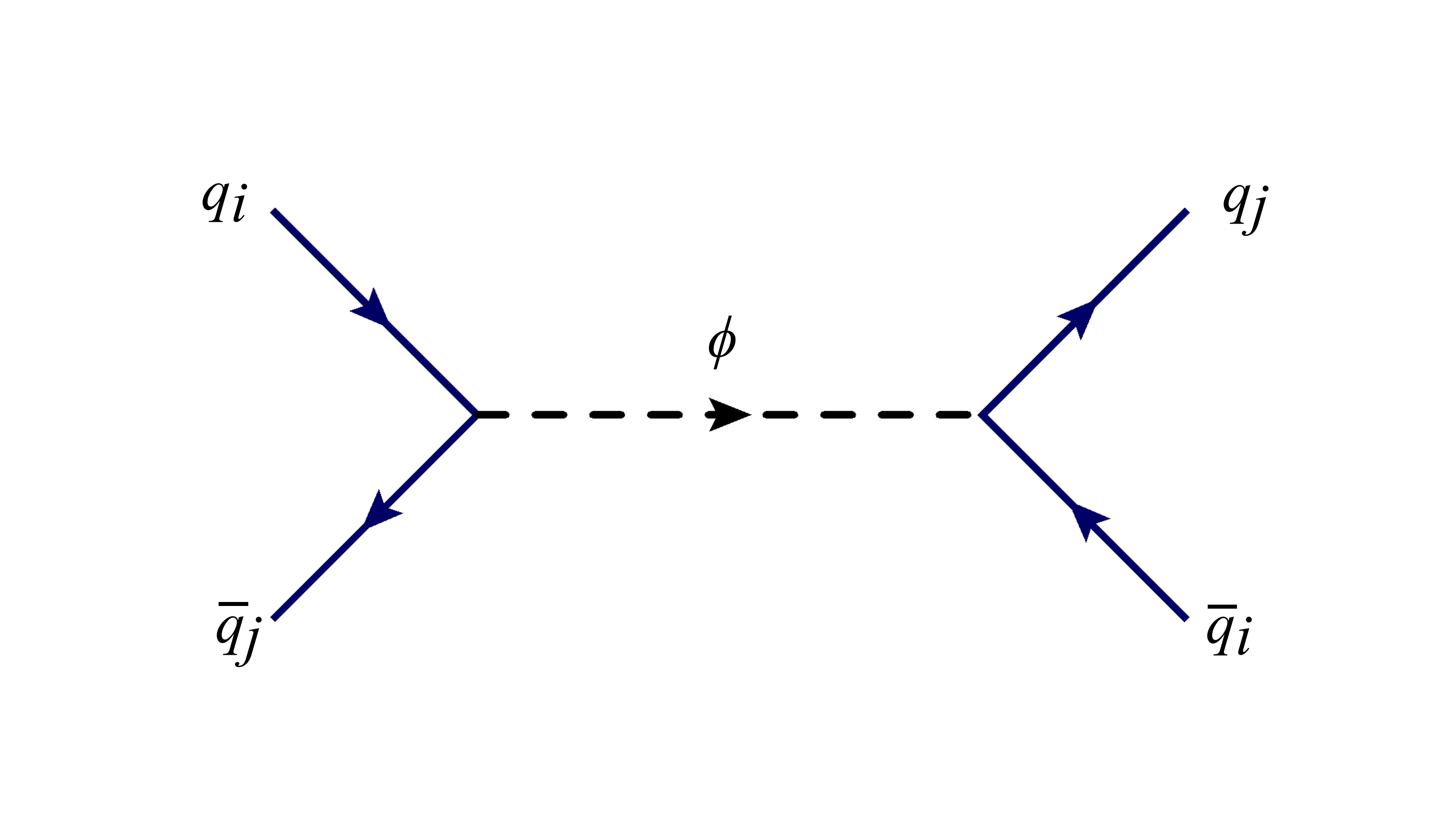}
		\caption{}
		\label{Fig.: FCNC quarks Higgs}
	\end{subfigure}
	\hfill
	\begin{subfigure}{0.5 \textwidth}
		\centering
		\includegraphics[scale=0.15]{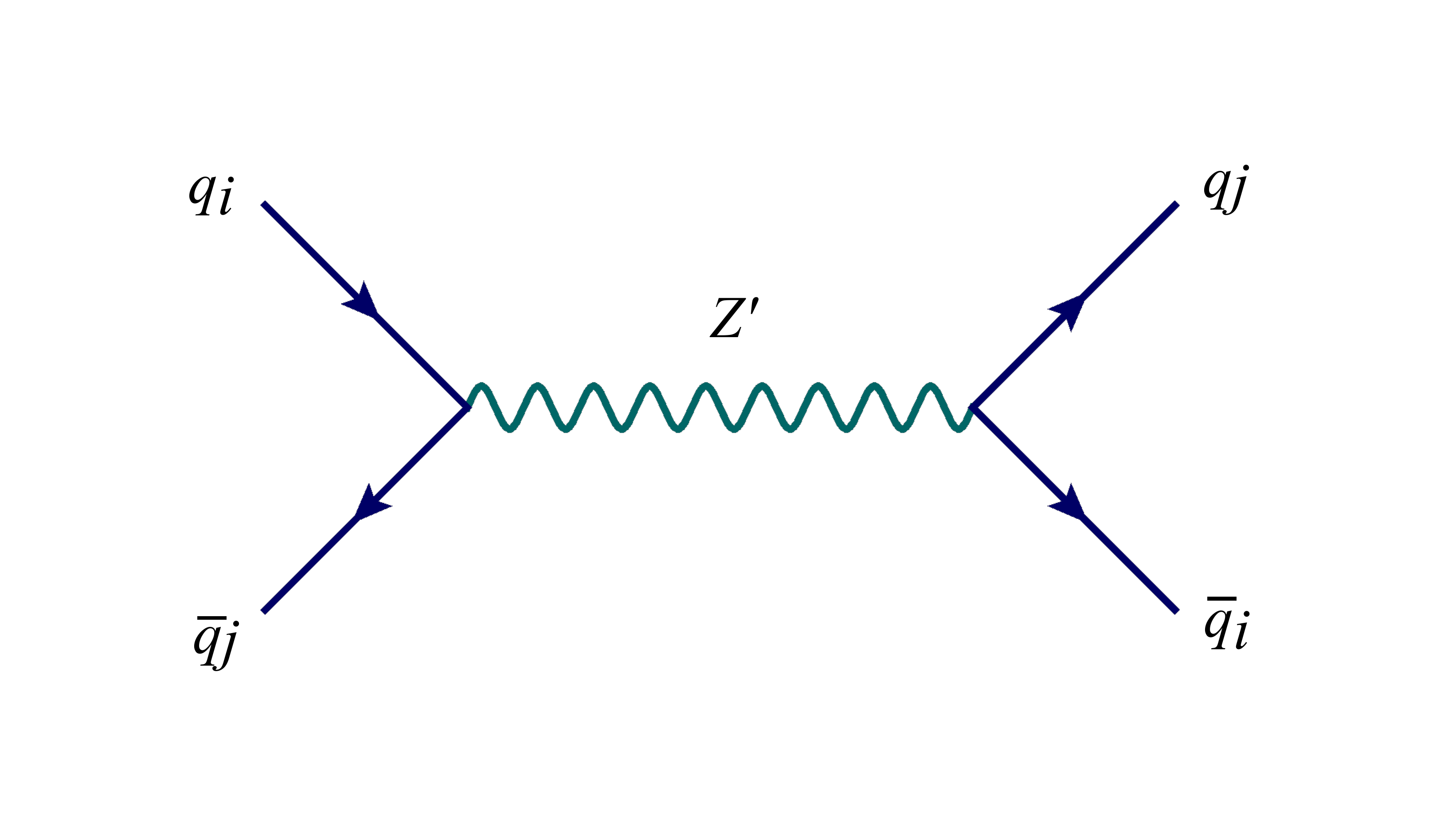}
		\caption{}
		\label{Fig.: FCNC quarks Z'}
	\end{subfigure}
	\caption{Tree level contribution to neutral meson mixing by \textbf{(a)} SM Higgs exchange and  \textbf{(b)} $Z'$ exchange.}
	\label{Fig: Neutral Meson Mixing}
\end{figure}  
Due to the tree level FCNCs that are present in our model, it is important to estimate the new physics effects on observables in the neutral meson mixing systems $D^0-\overline{D^0}$, $K^0-\overline{K^0}$, $B_d^0-\overline{B_d^0}$ and $B_s^0-\overline{B_s^0}$. 
The dominating contribution to these processes originates from three level $\phi$ and $Z'$ exchange as shown in Fig.~\ref{Fig: Neutral Meson Mixing} \footnote{In principle there are also tree level FCNCs mediated by the $Z$. However, these are rather small since the $Z$ couples diagonally in the interaction basis and small non-diagonal contributions arise only via mixing with the VL fermions.}.

In the following we apply an effective operator approach where contributions from heavy particles are integrated out. Then, the  Hamiltonian responsible for mixing in the neutral meson system reads 
\begin{equation}
\begin{split}
    \mathcal{H}_{\mathrm{eff}} =  &-\dfrac{1}{2 m_{\phi}^2} \left(\overline{\hat{q}}_i\left[ (\hat{\mathbf{Y}}_{q})_{ij}  \dfrac{1+\gamma_5}{2} + (\hat{\mathbf{Y}}^*_{q})_{ij}\dfrac{1-\gamma_5}{2} \right] \hat{q}_j \right)^2 \\
    &+ \dfrac{1}{2 M^2_{Z'}} \left( \overline{\hat{q}}_i \gamma_\mu \left[ \left[\hat{g}_L^{q}(Z')\right]_{ij}  \dfrac{1-\gamma_5}{2} + \left[\hat{g}_R^{q}(Z')\right]_{ij}\dfrac{1+\gamma_5}{2} \right] \hat{q}_j\right)^2 \, ,
    \end{split}
\end{equation}
where $\hat{q}_i$ and $\hat{q}_j$ indicate the quark fields participating in the mixing. Regarding a general meson $P$, the transition matrix element is defined by $ M_{12}^P = \langle P|\mathcal{H}_{\mathrm{eff}}|\overline{P}\rangle$ from which we can calculate the physical observable mass splitting 
$$
\Delta m_P = 2 \mathrm{Re}(M_{12}^P) \, 
$$
and in the case of the kaon the CP violation parameter  $|\epsilon_K |\simeq \mathrm{Im}(M_{12}^K)/(\sqrt{2} \Delta m_K )$.
For the relevant 4-fermion operators we use the hadronic transition matrix elements \cite{Ciuchini:1998ix}
\begin{equation}
\begin{split}
    \langle P|\overline{q}_i\dfrac{(1\pm \gamma_5)}{2} q_j \overline{q}_i \dfrac{(1\mp \gamma_5)}{2} q_j|\overline{P}\rangle &=  f_P^2 m_P \left( \dfrac{1}{24} +\dfrac{1}{4} \dfrac{m_P^2}{(m_{q_{i}}+ m_{q_{j}})^2 }\right) B_4 \, ,\\
    \langle P|\overline{q}_i \dfrac{(1\pm \gamma_5)}{2} q_j \overline{q}_i \dfrac{(1\pm \gamma_5)}{2} q_j|\overline{P}\rangle &=  - \dfrac{5}{24}f_P^2 m_P  \dfrac{m_P^2}{(m_{q_{i}}+ m_{q_{j}})^2 } B_2 \, ,\\
    \langle P|\overline{q}_i \gamma_\mu\dfrac{(1\pm \gamma_5)}{2} q_j  \overline{q}_i \gamma^\mu\dfrac{(1\pm \gamma_5)}{2} q_j|\overline{P}\rangle &=   \dfrac{1}{3}m_P f_P^2 B_1 \, ,
\end{split}
\end{equation}
which depend on the B parameters, the meson decay constant $f_P$ and the meson mass $m_P$.
With this at hand, the complete transition matrix element takes the form\footnote{We use here a Fierz rearrangement for the mixed left -and right-handed current of the $Z'$ (see\cite{Arnan:2019uhr}).
}
\begin{equation}
\begin{split}
\label{Eq.: Formula M12}
    M_{12}^P  &=  -\dfrac{f_P^2 m_P}{2 m_\phi^2}\Bigg[-\dfrac{5}{24}\dfrac{m_P^2}{(m_{q_{i}}+ m_{q_{j}})^2 } \left((\hat{\mathbf{Y}}_{q})^2_{ij}+ (\hat{\mathbf{Y}}^*_{q})^2_{ij} \right) \cdot B_2 \cdot \eta_2(\mu)  \\
    &+ (\hat{\mathbf{Y}}_{q})_{ij}(\hat{\mathbf{Y}}^*_{q})_{ij}\left( \dfrac{1}{12} + \dfrac{1}{2}\dfrac{m_P^2}{(m_{q_{i}}+ m_{q_{j}})^2 }\right) \cdot B_4 \cdot \eta_4(\mu)  \Bigg] \\
    &+ \dfrac{f_P^2 m_P}{2 M_{Z'}^2} \dfrac{1}{3} \left( \left[\hat{g}_L^{q}(Z')\right]^2_{ij} + \left[\hat{g}_R^{q}(Z')\right]^2_{ij} \right)\cdot B_1 \cdot \eta_1(\mu)\\
    &+ \dfrac{  f_P^2 m_P}{ M_{Z'}^2} \left[\hat{g}_L^{q}(Z')\right]_{ij} \left[\hat{g}_R^{q}(Z')\right]_{ij} \left(\dfrac{1}{12} + \dfrac{1}{2} \dfrac{m_P^2}{(m_{q_{i}}+ m_{q_{j}})^2 } \right) \cdot B_4 \cdot \eta_4(\mu)
    \, .
\end{split}
\end{equation}
 In above equation $\eta_1, ~\eta_2$ and $\eta_4$ are QCD correction factors for the  Wilson Coefficients (WC) that account for going from the heavy mass scale to the hadronic scale $\mu$. We explain their calculation in the following.
 
Generally, the effective $\Delta F = 2$ Hamiltonian is described by
\begin{equation}
    \mathcal{H}_{\mathrm{eff}}^{\Delta F = 2} = \sum_{i = 1}^5 C_i Q_i +  \sum_{i = 1}^3 \tilde{C}_i \tilde{Q}_i \, ,
\end{equation}
where $C_i$ are the WCs for a basis of 4-fermion operators
\begin{equation}
    \begin{split}
        Q_1 = \overline{q}_{iL}^\alpha \gamma_\mu &q_{jL}^\alpha \overline{q}_{iL}^\beta \gamma^\mu q_{jL}^\beta \, ,\indent   Q_2 = \overline{q}_{iR}^\alpha  q_{jL}^\alpha \overline{q}_{iR}^\beta  q_{jL}^\beta  ,\indent 
         Q_3 = \overline{q}_{iR}^\alpha  q_{jL}^\beta \overline{q}_{iR}^\beta  q_{jL}^\alpha \, , \\
          &Q_4 = \overline{q}_{iR}^\alpha  q_{jL}^\alpha \overline{q}_{iL}^\beta  q_{jR}^\beta \, , \indent \indent
           Q_5 = \overline{q}_{iR}^\alpha  q_{jL}^\beta \overline{q}_{iL}^\beta  q_{jR}^\alpha \, ,
    \end{split}
\end{equation}
and  $\tilde{Q_i}$ symbolizes the corresponding operator with interchanged $L \leftrightarrow R$.

Since the WCs are evaluated at the scale of new physics, they need to be evolved down to the hadronic energy scale for consistency.  Following \cite{Becirevic:2001jj},  the relation between the WC at the heavy scale $M_H$ and the WC at the scale $\mu$ is given by
\begin{equation}
\label{Eq.: Magic Number Equation}
    C_r(\mu) = \sum_{i} \sum_{s} (b_i^{(r,s)}+\eta c_i^{(r,s)}) \eta^{a_i} C_s(M_H) \, ,
\end{equation}
where $\eta = \alpha_s(M_H)/ \alpha_s(m_t)$ and the coefficients $b_i^{(r,s)}$, $ c_i^{(r,s)} $ and $a_i$ are  referred to as magic numbers.\\
For the $B_d$ and $B_s$ meson we take the magic numbers  from \cite{Becirevic:2001jj} and use the B-parameters $(B_1, B_2, B_4) = (0.87, 0.82, 1.16)$. The decay constants are given by $f_{B_d} =0.240 $ GeV and $f_{B_s} = 0.295$ GeV while we use $m_{B_d} = 5.281 $ GeV and $m_{B_s} = 5.370$ GeV for the neutral meson masses.  At the scale $M_H = m_\phi = 125.1$ GeV there are contributions from the operators $Q_2$ and $Q_4$. Using Eq.~(\ref{Eq.: Magic Number Equation}) with $\mu = m_b$ we find
\begin{equation}
\begin{split}
     C_2(\mu) &= 1.650 \cdot C_2 (M_H) \, , \indent C_3(\mu) = -0.014 \cdot C_2 (M_H) \, , \\
     C_4(\mu) &= 2.259 \cdot C_4 (M_H) \, , \indent C_5(\mu) = 0.056 \cdot C_4 (M_H) \, .
    \end{split}
\end{equation}
Although the operators $Q_3$ and $Q_5$ are induced via operator mixing, their contribution is negligibly small and for the correction factors we find $\eta_2 (\mu) = 1.650$ and $\eta_4 (\mu)= 2.259$. Similarly, we proceed in the case of the $Z'$. Here, the operators $Q_1$ and $Q_4$ contribute at $M_H = M_{Z'} = 300$ TeV and we obtain
\begin{equation}
\begin{split}
    C_1(\mu) = 0.713 \cdot C_1 (M_H)\, , ~
     C_4(\mu) &= 5.446 \cdot C_4 (M_H) \, , ~ C_5(\mu) = 0.165 \cdot C_4 (M_H) \, .
\end{split}
\end{equation}
This yields $\eta_1(\mu)= 0.713$, $\eta_4(\mu) = 5.446$ and  a negligible contribution of the induced $Q_5$.\\
For the $K^0-\overline{K}^0$ system we use the magic numbers  from \cite{Ciuchini:1998ix},   $(B_1, B_2, B_4) = (0.60, 0.66, 1.03)$, $f_K =0.160 $ GeV and $m_K= 0.498$ GeV. At the scale $\mu = 2 $ GeV the WC induced by the Higgs effective operators are
\begin{equation}
    \begin{split}
         C_2(\mu) &= 2.210 \cdot C_2 (M_H) \, , \indent C_3(\mu) = 0.003 \cdot C_2 (M_H) \, , \\
     C_4(\mu) &= 3.523 \cdot C_4 (M_H) \, , \indent C_5(\mu) = 0.1289 \cdot C_4 (M_H) \, ,
    \end{split}
\end{equation}
which results in $\eta_2(\mu)= 2.210$ and  $\eta_4(\mu) = 3.523$. For the effective operators generated by the $Z'$ we find 
\begin{equation}
\begin{split}
    C_1(\mu) =  0.674\cdot C_1 (M_H)\, , ~
     C_4(\mu) &=  8.181\cdot C_4 (M_H) \, , ~ C_5(\mu) =  0.329\cdot C_4 (M_H) \, ,
\end{split}
\end{equation}
and hence $\eta_1(\mu)= 0.674$ and $\eta_4(\mu) = 8.181$.\\
Finally, for the $D$ meson the B-parameters are given by $(B_1, B_2, B_4) = (0.865, 0.82, 1.08)$. The decay constant and meson mass are given by $f_D =0.200 $ GeV and $m_D= 1.864$ GeV, respectively. Using the magic numbers from \cite{UTfit:2007eik} to evolve the WCs induced by the Higgs down to $\mu = 2.8 $ GeV we find
\begin{equation}
    \begin{split}
         C_2(\mu) &= 1.906 \cdot C_2 (M_H) \, , \indent C_3(\mu) =  -0.006\cdot C_2 (M_H) \, , \\
     C_4(\mu) &=  2.903\cdot C_4 (M_H) \, , \indent C_5(\mu) =  0.097\cdot C_4 (M_H) \, ,
    \end{split}
\end{equation}
Thus, $\eta_2(\mu)= 1.906$ and $\eta_4(\mu) = 2.903$. For the $Z'$ similar evaluation as before reveals
\begin{equation}
\begin{split}
    C_1(\mu) =  0.690\cdot C_1 (M_H)\, , ~
     C_4(\mu) &=  6.939\cdot C_4 (M_H) \, , ~ C_5(\mu) =  0.263\cdot C_4 (M_H) \, ,
\end{split}
\end{equation}
and therefore $\eta_1(\mu)= 0.690$ and $\eta_4(\mu) = 6.939$.
 Evidently, in all cases the induced  operators play a subdominant role and  therefore we will not consider them in our analysis.
 
Since the model reproduces the correct fermion masses and CKM mixing angles, contributions of the usual box diagrams to the neutral meson mixing are almost unaltered compared to the SM case\footnote{Although there are further diagrams with an intermediate VL fermion in the loop, these contributions are highly suppressed.  The VL fermions are weak singlets and a tiny coupling to $W^\pm$ is only induced by fermion mass mixing making these small deviations negligible.}. Due to chirality, the new contribution of the Higgs exchange diagram cannot interfere with the purely left-handed SM contribution. Furthermore, the leading contribution from the $Z'$ exchange originates from the operator with mixed left -and right-handed current. Thus, we will not consider any interference effects here and write the  total mass difference as
\begin{equation}
    \Delta m_P^{\mathrm{tot}} = \Delta m_P^{\mathrm{SM}} + \Delta m_P^{\mathrm{NP}}
\end{equation}
Within the theoretical and experimental uncertainties $\Delta m_P^{\mathrm{tot}}$ should agree with the measured values $\Delta m_P^{\mathrm{exp}}$. Therefore we demand the new physics contribution to agree within $3\sigma$ with  the difference $ \Delta m_P^{\mathrm{exp}}-\Delta m_P^{\mathrm{SM}}$.

The Standard Model prediction for the $B^0$-meson systems are $\Delta m_{B_d}^{\mathrm{SM}} = (3.475 \pm 0.513)\times 10^{-13}$ GeV and $\Delta m_{B_s}^{\mathrm{SM}} = (1.205 \pm 0.178)\times 10^{-11}$ GeV \cite{Artuso:2015swg}, while the measured values are $\Delta m_{B_d}^{\mathrm{exp}} = (3.334 \pm 0.013)\times 10^{-13}$ GeV and $\Delta m_{B_s}^{\mathrm{exp}} = (1.169 \pm 0.001)\times 10^{-11}$ GeV \cite{Zyla:2020zbs} which leads to
\begin{equation}
    \begin{split}
        \Delta m_{B_{d}}^{\mathrm{exp}}-\Delta m_{B_{d}}^{\mathrm{SM}} = (-0.141 \pm 0.513)\times 10^{-13} \, \mathrm{GeV} \, ,\\
        \Delta m_{B_{s}}^{\mathrm{exp}}-\Delta m_{B_{s}}^{\mathrm{SM}} = (-0.036 \pm 0.178)\times 10^{-11} \, \mathrm{GeV} \, .
    \end{split}
\end{equation}
For the $K^0$-meson system the theoretical prediction of the short distance contribution is $\Delta m_{K}^{\mathrm{SM}} = 3.074 \times 10^{-15}$ GeV \cite{Buras:2015kwd}, where we will assume a $30\%$ uncertainty due to unknown large distance contributions that cannot be calculated from first principles \cite{NguyenTuan:2020xls}. Experiments measured $\Delta m_{K}^{\mathrm{exp}} = (3.484 \pm 0.006)\times 10^{-15}$ GeV \cite{Zyla:2020zbs} which gives
\begin{equation}
    \Delta m_{K}^{\mathrm{exp}}-\Delta m_{K}^{\mathrm{SM}} = (0.410 \pm 0.922)\times 10^{-15} \, \mathrm{GeV} \, .
\end{equation}
Since we assumed only real Yukawa couplings in the down quark sector we do not get any additional contribution to $|\epsilon_K|$. Finally, the mass difference for neutral D-mesons is measured to be  $\Delta m_{D}^{\mathrm{exp}} =(6.253^{+2.699}_{-2.896} )\times 10^{-15}$ GeV \cite{Zyla:2020zbs}. However, just as it is the case for $K^0$-mesons, the theoretical prediction of the Standard Model contribution to neutral $D$-meson mixing is subject to large uncertainties \cite{Golowich:2007ka}. We will therefore only demand the new physics contribution to be less than the  error on $\Delta m_{D}^{\mathrm{exp}}$.

Our results for the two benchmark points are given in Table~\ref{Tab.: FCNC observables in Quark Sector} and  agree  with the demanded limits. The additional contribution to $B_d^0-\overline{B_d^0}$ mixing is closest to experimental limits and could therefore provide an excellent channel to observe new physics effects in upcoming experiments.
\begin{table}[htb!]
\centering
\begin{tabular}{|c|c|c|}
\hline\hline
\multirow{2}{*}{\begin{tabular}[c]{@{}c@{}}\bf{Observable}\\ (in GeV)\end{tabular}} & \multicolumn{2}{c|}{\bf{Model Prediction}}                                          \\ \cline{2-3} 
                                                                               & \multicolumn{1}{c|}{\bf{BP1}}   & \bf{BP2}   \\ \hline \hline
$\Delta m^{\mathrm{NP}}_{B_d}$                                                               & \multicolumn{1}{c|}{$-1.402 \times 10^{-13}$ } & $-1.495 \times 10^{-14}$  \\ \hline
$\Delta m^{\mathrm{NP}}_{B_s}$                                                               & \multicolumn{1}{c|}{$2.663 \times 10^{-14}$ } & $3.003 \times 10^{-14}$  \\ \hline
$\Delta m^{\mathrm{NP}}_{D}$                                                                 & \multicolumn{1}{c|}{$2.405 \times 10^{-15}$ } & $2.036 \times 10^{-15}$  \\ \hline
$\Delta m^{\mathrm{NP}}_{K}$                                                                 & \multicolumn{1}{c|}{$0.504 \times 10^{-15}$ } & $0.109 \times 10^{-15}$  \\ \hline
\end{tabular}
\caption{New physics contribution to observable in  neutral meson mixing  for the two given benchmark points. }
\label{Tab.: FCNC observables in Quark Sector}
\end{table}

\subsection{Charged Lepton Flavor Violation}
\begin{figure}[htb!]
	\begin{subfigure}{0.5 \textwidth}
		\centering
		\includegraphics[scale=0.15]{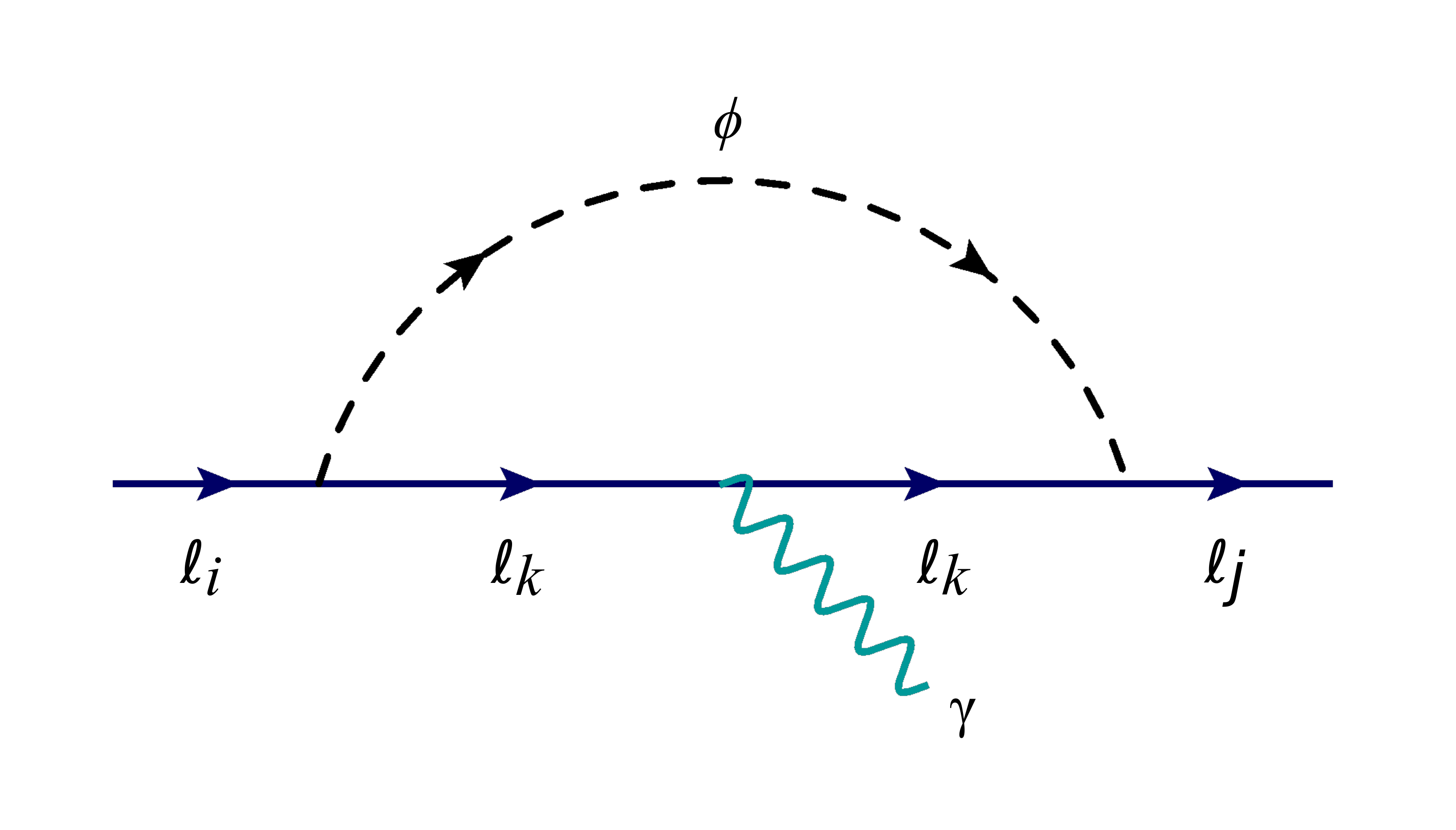}
		\caption{}
		\label{Fig.: Mu to E Gamma Higgs}
	\end{subfigure}
	\hfill
	\begin{subfigure}{0.5 \textwidth}
		\centering
		\includegraphics[scale=0.15]{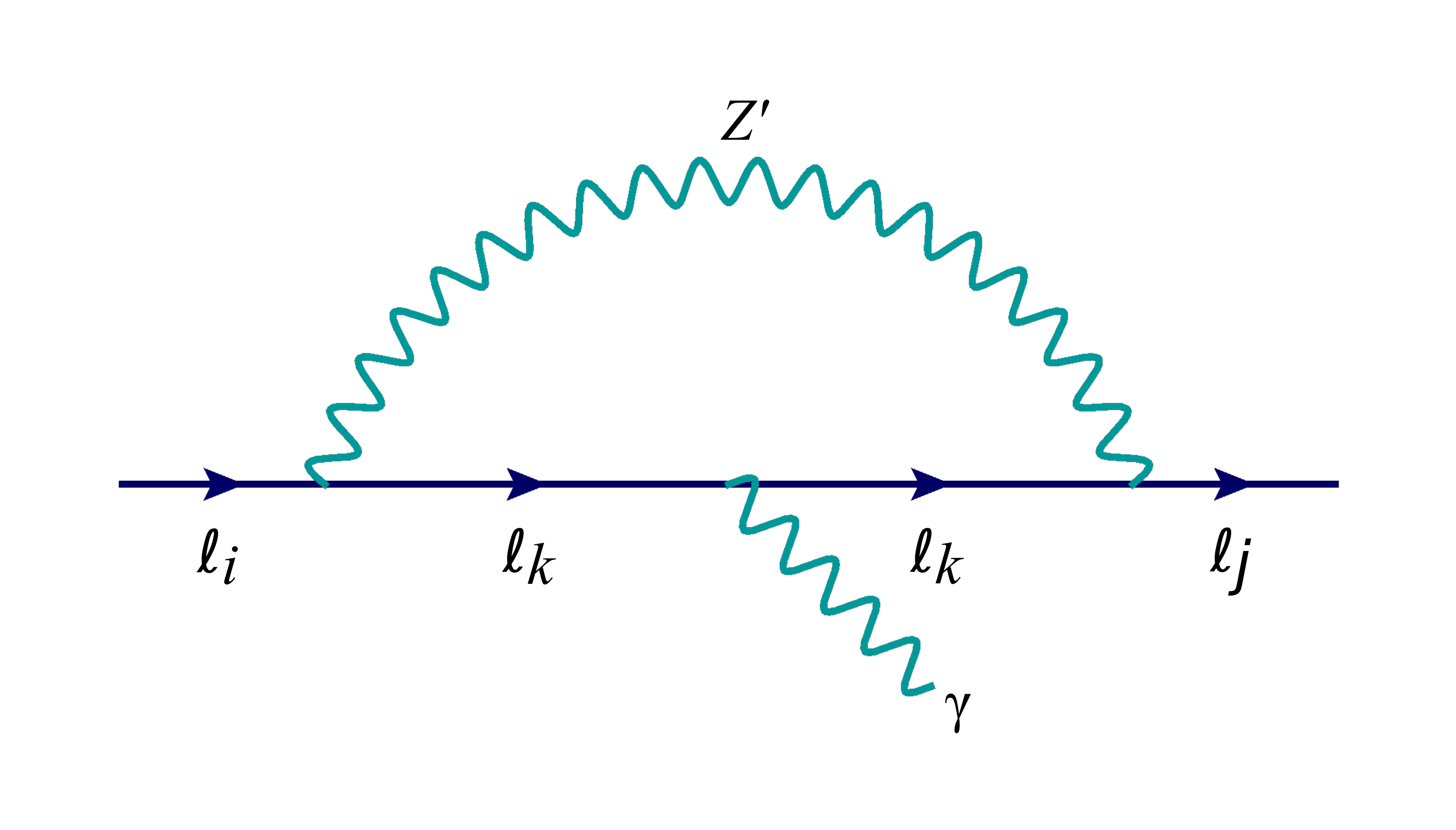}
		\caption{}
		\label{Abb.:  Mu to E Gamma Z'}
	\end{subfigure}
	\caption{One-loop contributions to the decay $\ell_i \rightarrow \ell_j \gamma$ via \textbf{(a)} SM Higgs exchange and  \textbf{(b)} $Z'$ exchange.}
	\label{Fig.: LFV Diagrams}
\end{figure}
The flavor violating couplings of the Higgs and the $Z'$ equally impact observable in the charged lepton sector. Considering the  process $\ell_i \rightarrow \ell_j \gamma$, there are two diagrams contributing at one-loop to the decay and we illustrate them in Fig.~\ref{Fig.: LFV Diagrams}. The contribution from the one-loop diagram mediated by the Higgs $\phi$ is given by \cite{Lavoura:2003xp}
\begin{equation}
\label{Eq.: decay width li to lj gamma}
    \Gamma(\ell_i \rightarrow \ell_j \gamma) = \sum_{k} \dfrac{(m_i^2-m_j^2)^3\left(|\sigma_L|^2+ |\sigma_R|^2\right)}{16 \pi m_i^3}
\end{equation}
with 
\begin{equation}
    \sigma_L = \dfrac{i Q_k}{16 \pi^2 m_\phi^2} \left[ (\rho  m_i+\lambda  m_j) \mathcal{F}_1(t)+ \nu m_k \mathcal{F}_2(t) \right] \, 
\end{equation}

\begin{equation}
    \sigma_R = \dfrac{i Q_k}{16 \pi^2 m_\phi^2} \left[ (\lambda  m_i+\rho  m_j) \mathcal{F}_1(t)+ \zeta m_k \mathcal{F}_2(t) \right] \, 
\end{equation}
and we sum over all internal fermions $\ell_k$ with electric charge $Q_k$.
The functions $\mathcal{F}_1(t)$  and $\mathcal{F}_2(t)$ are defined by
\begin{equation}
    \begin{split}
        \mathcal{F}_1(t) &= \dfrac{t^2-5t-2}{12(t-1)^3} + \dfrac{t\ln{t}}{2(t-1)^4} \, ,\\
        \mathcal{F}_2(t) &= \dfrac{t-3}{2(t-1)^2} + \dfrac{\ln{t}}{(t-1)^3} \, ,
    \end{split}
\end{equation}
with $t = m_k^2/m_\phi^2$ and the couplings 
$\rho = (\mathbf{\hat{Y}}_e)^*_{kj} (\mathbf{\hat{Y}}_e)_{ki} $,  $\lambda = (\mathbf{\hat{Y}}_e)_{jk} (\mathbf{\hat{Y}}_e)^*_{ik}  $,  $\nu = (\mathbf{\hat{Y}}_e)^*_{kj} (\mathbf{\hat{Y}}_e)^*_{ik} $ and $\zeta = (\mathbf{\hat{Y}}_e)_{jk} (\mathbf{\hat{Y}}_e)_{ki}  $.
For the contribution of the $Z'$ exchange  Eq.~(\ref{Eq.: decay width li to lj gamma}) equally applies but with the replacement
\begin{equation}
    \sigma_L = \dfrac{i Q_k}{16 \pi^2 M_{Z'}^2} \left[ (\rho'  m_i+\lambda'  m_j) \mathcal{F}_3(t)+ \nu' m_k \mathcal{F}_4(t) - \zeta'\dfrac{m_i m_j m_k}{M^2_{Z'}} \mathcal{F}_5(t) \right] \, 
\end{equation}

\begin{equation}
    \sigma_L = \dfrac{i Q_k}{16 \pi^2 M_{Z'}^2} \left[ (\lambda'  m_i+\rho'  m_j) \mathcal{F}_3(t)+ \zeta' m_k \mathcal{F}_4(t) - \nu'\dfrac{m_i m_j m_k}{M^2_{Z'}} \mathcal{F}_5(t) \right] \, 
\end{equation}
where $t = m_k^2/ M_{Z'}^2$ and we defined the functions
\begin{equation}
    \begin{split}
        \mathcal{F}_3(t) &= \dfrac{-5t^3+9t^2-30t+8}{12(t-1)^3} + \dfrac{3t^2\ln{t}}{2(t-1)^4} \, ,\\
        \mathcal{F}_4(t) &= \dfrac{t^2+t+4}{2(t-1)^2} - \dfrac{3t\ln{t}}{(t-1)^3} \, , \\
        \mathcal{F}_5(t) &= \dfrac{-2t^2+7t-11}{6(t-1)^3} + \dfrac{\ln{t}}{(t-1)^4} \, .
    \end{split}
\end{equation}
The couplings in this notation are given by $\lambda' = \left[ \hat{g}_L^\ell(Z') \right]^*_{kj}\left[ \hat{g}_L^\ell(Z') \right]_{ki}$, $\rho' = \left[ \hat{g}_R^\ell(Z') \right]^*_{kj}\left[ \hat{g}_R^\ell(Z') \right]_{ki}$, $\zeta' = \left[ \hat{g}_L^\ell(Z') \right]^*_{kj}\left[ \hat{g}_R^\ell(Z') \right]_{ki}$ and $\nu' = \left[ \hat{g}_R^\ell(Z') \right]^*_{kj}\left[ \hat{g}_L^\ell(Z') \right]_{ki}$.

\begin{figure}[htb!]
	\centering
	\includegraphics[scale=0.15]{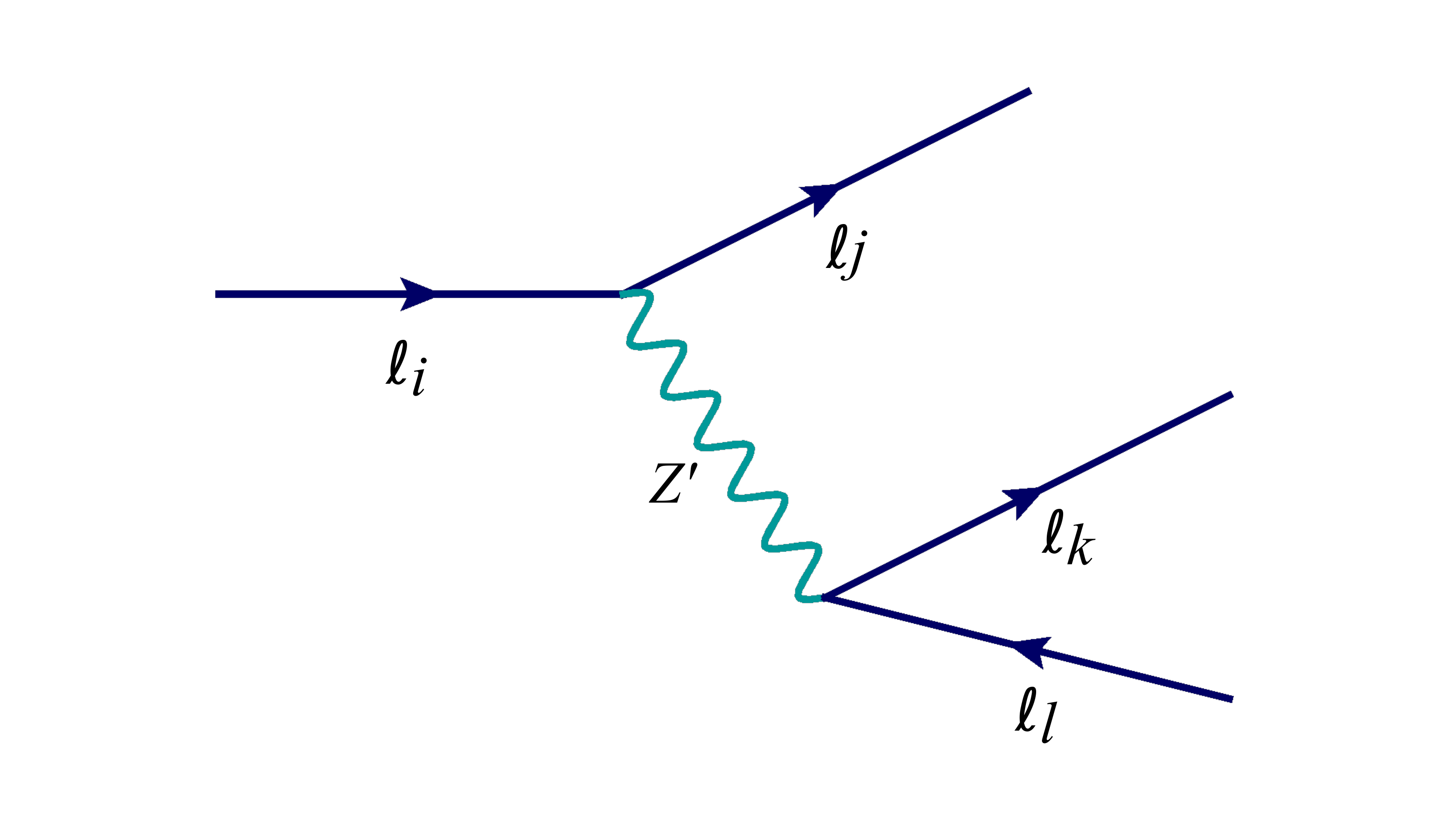}
	\caption{Tree level contribution of $Z'$ mediated tri-lepton decay. }
	\label{Fig.: Three lepton Decay}
\end{figure}

In addition, the flavor non-diagonal couplings of the $Z'$ and $\phi$ give rise to tree level contributions to the three lepton decay $\ell_i \rightarrow \overline{\ell}_j \ell_k \ell_l$. The partial decay width mediated by  $\phi$ exchange is given by
\begin{equation}
    \Gamma(\ell_i \rightarrow \overline{\ell}_j \ell_k \ell_l) = \dfrac{1}{1536 \pi ^3} \dfrac{m_i^5}{m_\phi^4} S |(\mathbf{\hat{Y}}_e)^*_{ij} (\mathbf{\hat{Y}}_e)_{kl}|^2 \, ,
\end{equation}
with a factor $S = 1$ for two equally charged fermions in the final state $(k = l)$ and $S =2 $ for $k \neq l$. The contribution from tree level  $Z'$ exchange (see Fig.~\ref{Fig.: Three lepton Decay}) can be written like
\begin{equation}
    \Gamma(\ell_i \rightarrow \overline{\ell}_j \ell_k \ell_l) = \dfrac{1}{1536 \pi ^3} \dfrac{m_i^5}{M_{Z'}^4}S \left[\dfrac{2}{S} |C_{LL}|^2+ \dfrac{2}{S} |C_{RR}|^2+  |C_{RL}|^2+ |C_{LR}|^2\right]^{ij}_{kl}
\end{equation}
where we use the abbreviation
\begin{equation}
   \left[ C_{ X Y}\right]^{ij}_{kl} = [\hat{g}^{\ell}_X(Z')]_{ij} [\hat{g}^{\ell}_X(Z')]_{kl}  \, ,
\end{equation}
with $X,Y \in \{L,R\}$ and $\hat{g}^{\ell}_X(Z')$ denoting the coupling matrix of the $Z'$ to charged leptons with chirality $X$ \cite{Aebischer:2019blw}. 
The additional factor two for the coefficients $C_{LL}$ and $C_{RR}$ arises due to the  fact that diagrams which differ by the replacement of two identical leptons can interfere with each other if the two currents are of same chirality.
From our analysis we find that the $Z' $ contribution to this process is much  larger than that of the  Higgs. The reason behind this is  that the Higgs boson has not only suppressed flavor non-diagonal couplings but also small flavor diagonal couplings to first and second generation leptons. 

From the total decay widths of the muon and tau lepton,  $\Gamma_\mu^\mathrm{tot} = 3.00\times10^{-19} $GeV and $\Gamma_\tau^\mathrm{tot} = 2.27\times10^{-12} $GeV \cite{Zyla:2020zbs}, the branching ratios for the decays $\mu^- \rightarrow e^- \gamma$,  $\tau^- \rightarrow e^- \gamma$, $\tau^- \rightarrow \mu^- \gamma$,  $\tau^- \rightarrow \mu^- e^+ e^-$, $\tau^- \rightarrow e^- e^+ e^-$ and $\mu^- \rightarrow e^- e^+ e^-$ are calculated for the two benchmark points  and can be seen in Table~\ref{Tab.: LFV decays model prediction}. In this scenario, all expected branching ratios are below the current experimental limits. However, the experimental sensitivity for the process $\mu^- \rightarrow e^- \gamma$ is not far from our model prediction and may be tested in future experiments. Finally, we emphasize the interesting fact that these flavor violating observable are approximately equal for each generation.

\begin{table}[ht!]
\centering
\begin{tabular}{|c|c|cc|}
\hline \hline
\multirow{2}{*}{\bf{Process}}           & \multirow{2}{*}{\bf{Experimental Limit}} & \multicolumn{2}{c|}{\bf{Model Prediction}}                          \\ \cline{3-4} 
                                         &                                           & \multicolumn{1}{c|}{\bf{BP1}}                    & \bf{BP2 }                   \\ \hline \hline
$\mathrm{BR}(\mu^- \rightarrow e^- \, \gamma)$    & $< 4.2 \times 10^{-13}$                   & \multicolumn{1}{c|}{$ 1.8 \times 10^{-14}$} & $ 6.3 \times 10^{-15}$ \\ \hline
$\mathrm{BR}(\tau^- \rightarrow e^- \, \gamma)$   & $< 3.3 \times 10^{-8}$                    & \multicolumn{1}{c|}{$ 2.2 \times 10^{-14}$} & $ 2.2 \times 10^{-14}$ \\ \hline
$\mathrm{BR}(\tau^- \rightarrow \mu^- \, \gamma)$ & $< 4.4 \times 10^{-8}$                    & \multicolumn{1}{c|}{$ 6.8 \times 10^{-15}$} & $ 3.0 \times 10^{-15}$ \\ \hline
$\mathrm{BR}(\mu^- \rightarrow e^-  e^+  e^-)$    & $< 1.0 \times 10^{-12}$                   & \multicolumn{1}{c|}{$1.3 \times 10^{-18}$}  & $ 3.9 \times 10^{-19}$ \\ \hline
$\mathrm{BR}(\tau^- \rightarrow e^- e^+  e^-)$    & $< 2.7 \times 10^{-8}$                    & \multicolumn{1}{c|}{$ 2.2 \times 10^{-17}$} & $1.8 \times 10^{-17}$  \\ \hline
$\mathrm{BR}(\tau^- \rightarrow \mu^- e^+  e^-) $ & $< 1.8 \times 10^{-8}$                    & \multicolumn{1}{c|}{$9.5 \times 10^{-18}$}  & $4.1 \times 10^{-18}$  \\ \hline \hline
\end{tabular}
\caption{Predicted branching ratios for certain LFV processes and current experimental limits \cite{Zyla:2020zbs}. }
\label{Tab.: LFV decays model prediction}
\end{table}

\subsection{Other Implications}\label{SEC-05}
The presence of a heavy neutral gauge boson and heavy vector-like fermions in our framework will have a plethora of implications at collider experiments.  Throughout our analysis, we choose $Z'$ mass scale of $\mathcal{O}(100)$ TeV, although the vector-like fermions might exist at a much lower scale (a few TeV), accessible to the future collider experiments. Note that the mass scales of these new heavy particles can be accommodated at a much lower scale with proper adjustment of benchmark parameters and which is beyond the scope of this study. However, we will briefly mention the collider implications and testable consequences on flavor anomalies for the low-scale realization.  After being pair produced through the $s-$ channel $Z/\gamma $ exchange at the $pp$ collider, singly charged vector-like leptons will lead to $2l +{E\!\!\!\!/}_{T}, 4l +{E\!\!\!\!/}_{T}$ signatures and vector-like quarks will lead to promising $jj+4l, b\Bar{b}+4l, t\Bar{t}+4l$ signatures at the LHC.  There has been an extensive study of vector-like lepton searches at future collider experiments (such as HL-LHC, HE-LHC, and FCC-hh), and it has been shown that a $SU(2)_L$ singlet vector-like lepton can be probed up to a mass of  $\sim 3$ TeV at $100$ TeV future collider looking at multi-leptons, and ${E\!\!\!\!/}_{T}$ final state signature \cite{Bhattiprolu:2019vdu}. Moreover, Ref.~\cite{Chala:2018qdf} has studied the discovery prospects of vector-like quarks looking at multi-lepton final state signatures (in association with jets or $t/b$ quarks) at future collider experiments, and a mass up to $\sim 7$ TeV vector-like quark can be probed at $100$ TeV collider. Test of these vector-like families will be a good test for our model at future collider experiments. It is worth noting that there has been an increase in attention since there are various flavor anomalies in the muon sectors, such as  muon $g-2$ \cite{Muong-2:2021ojo}, $R_{K}^{(*)}$ \cite{Aaij:2021vac, Aaij:2017vbb, LHCb:2019hip, Belle:2019gij} anomalies. We find that it is very difficult to address the $R_{K}^{(*)}$ \cite{Aaij:2021vac, Aaij:2017vbb, LHCb:2019hip, Belle:2019gij} anomalies at tree level in our framework since it suffers from strong flavor constraints from $B_S - \Bar{B}_S$ mixing. However, it is worth noting that muon $g-2$ anomaly can be easily addressed with correct strength and sign in our framework due to chiral enhancement from vector-like family for heavy vector-like fermions inside the loop \footnote{In Ref.~\cite{Dermisek:2020cod}, it was demonstrated in a different framework that the muon $g-2$ anomaly can be addressed with a 45 TeV of vector-like lepton within the loop.}. However, it is worth mentioning that it suffers from fine-tuning in the absence of additional symmetries since this results in substantial muon mass corrections at loop level. We should note that the muon $g-2$ contributions from the two BPs are three orders of magnitude lower than the experimental finding and muon mass correction is under control. 
Although the mass of the neutral heavy gauge boson $Z'$ in our theory is on the order of $\mathcal{O}(100)$ TeV, complementary probe from two body scattering might be found in a future muon collider experiment. It has recently been demonstrated that a $100$ (400) TeV $Z'$ scenario can be probed at a future muon collider experiment with a center of mass energy of  $\sqrt{s}=3$ TeV and an integrated luminosity of $\mathcal{L}=$ 1 ab$^{-1}$ looking at $\mu^+ \mu^- \to \mu^+ \mu^-$ signature while considering the strength of the gauge coupling $g_X=1 ~(\sqrt{4 \pi})$ \cite{Huang:2021biu, Huang:2021nkl}. Note that there will be additional contributions to the  Higgs observable as well. However, these contributions will be negligible due to new physics at moderately higher scales as mentioned in the BPs. For example,  our framework leads to SM Higgs mediated flavor violating signals like $h \to e \tau, \mu \tau$ which can be a clear sign of new physics. We analyze and find that the strength of these Higgs-mediated flavor violating signals is many orders of magnitude lower than the present experimental sensitivity. 
\section{Conclusions }\label{SEC-06}
We propose a \textit{flavor seesaw} mechanism where large hierarchies and patterns of mixings of quarks and leptons arise naturally. The mechanism requires two additional vector generations of fermions with TeV-ish masses, which lead to a seesaw-like fermion mass matrices with tree level and loop contributions. The masses of the third and second generations of quarks and leptons arises at tree level, while gauge boson mediated loop corrections generate the masses of the first generation. We realize the \textit{flavor seesaw} mechanism in a model where the SM is extended by an extra $U(1)_{X}$ symmetry. Our fermion mass generation mechanism is universally applicable to the up-type quarks, down-type quarks, charged leptons, and neutrinos. 

Counting parameters one can see that the model does not predict masses. Instead it maps the observed hierarchies and patterns of SM fermion masses onto the extended theory with Yukawa couplings $\mathcal{O}(1)$. An impressive 13 orders of magnitude of the SM from the light active neutrino mass scale to the top quark mass becomes thus much simpler. Note that the \textit{flavor seesaw} mechanism does not require any additional ingredient such as additional discrete flavor symmetries or multiple scalars. The mechanism automatically creates the exhibited patterns after diagonalization and it is therefore quite appealing to explain the hierarchical mass pattern and mixing regularities of the SM in such a manner. For our model we explicitly show how the mapping works numerically for $M_Z' = 300$ TeV.

The model has a plethora of implications and testable effects. Among these, the most promising one is the inescapable and complementary flavor-violating signal in upcoming experiments. We have analyzed all these flavor violating signals in the quark as well as lepton sector and find that $l_i \to l_j \gamma $ process and  $B_d^0-\overline{B_d^0}$ mixing is closest to current experimental limits and could therefore provide an excellent signal to observe new physics effects in upcoming experiments. The presence of heavy vector-like fermions at a few TeV scale in our model will result in promising multi-lepton signals associated with jets or missing energy, which can be probed in future collider experiments like FCC-hh. Note that the discussed benchmark point does not imply that other solutions with lower $M_Z'$ could not exist. Such solutions would lead to further phenomenological consequences and their inter-dependence could be an interesting method to test the model further. For instance, due to chiral enhancement inside the loop, a relatively light $Z'$ can potentially address the observed discrepancy in muon $g-2$ measurement. Another interesting feature is that lepton number is  conserved such that neutrinos have naturally only Dirac masses. 

The seesaw flavor mechanism can be implemented in different ways. We demonstrated the mechanism for an extra $U(1)_{X}$ symmetry, but other solutions should exist. A generalization of the \textit{flavor seesaw} mechanism to non-abelian theories seems also possible and embeddings into GUT groups appear also possible. This will be studied in a forthcoming paper.

\bibliographystyle{utphys}
\bibliography{reference}
\end{document}